\documentclass[aps,prl,twocolumn,superscriptaddress]{revtex4-2}
\usepackage{amsmath}
\usepackage{mathrsfs}
\usepackage{dsfont}
\usepackage{amssymb}
\usepackage{textcomp}
\usepackage{hyperref}
\usepackage{graphicx}
\usepackage{float}
\usepackage{siunitx}
\usepackage{color}
\usepackage[dvipsnames]{xcolor}
\usepackage{bm}
\usepackage{soul}
\usepackage{comment}
\usepackage{ulem}

\usepackage{mathtools}
\usepackage{appendix}
\usepackage{caption}
\usepackage{subcaption}
\usepackage{booktabs}
\usepackage{array}
\usepackage[justification=raggedright]{caption}

\makeatletter
\def\@seccntformat#1{\csname the#1\endcsname.\quad}

\makeatother

\begin{document}
	
	\title{Orbital angular momentum radiation and polarization of relativistic electrons in magnetic field}
	
	\author{Ziqiang Huang}
	%\email{huangzq69@mail2.sysu.edu.cn}
	\affiliation{Sino-French Institute of Nuclear Engineering and Technology, Sun Yat-Sen University, Zhuhai 519082, China}
	\author{Qi Meng}
	%\email{mengq8@mail2.sysu.edu.cn}
	\affiliation{Sino-French Institute of Nuclear Engineering and Technology, Sun Yat-Sen University, Zhuhai 519082, China}
	\author{Xuan Liu}
	%\email{liux666@mail2.sysu.edu.cn}
	\affiliation{Sino-French Institute of Nuclear Engineering and Technology, Sun Yat-Sen University, Zhuhai 519082, China}
	\author{Wei Ma}
	%\email{mawei25@mail.sysu.edu.cn}
	\affiliation{Sino-French Institute of Nuclear Engineering and Technology, Sun Yat-Sen University, Zhuhai 519082, China}
	\author{Zhen Yang}
	%\email{yangzh97@mail.sysu.edu.cn}
	\affiliation{Sino-French Institute of Nuclear Engineering and Technology, Sun Yat-Sen University, Zhuhai 519082, China}
	\author{Liang Lu}
	%\email{luliang3@mail.sysu.edu.cn}
	\affiliation{Sino-French Institute of Nuclear Engineering and Technology, Sun Yat-Sen University, Zhuhai 519082, China}
	%\affiliation{United Laboratory of Frontier Radiotherapy Technology of Sun Yat-sen University \& Chinese Academy of Sciences Ion Medical Technology Co., Lid., Guangzhou 510000, China}
	\author{Alexander J. Silenko}
	%\email{alsilenko@mail.ru}
	\affiliation{Bogoliubov Laboratory of Theoretical Physics, Joint Institute for Nuclear Research, Dubna 141980, Russia}
	\author{Pengming Zhang}
	%\email{zhangpm5@mail.sysu.edu.cn}
	\affiliation{School of Physics and Astronomy, Sun Yat-sen University, Zhuhai 519082, China}
	\author{Liping Zou}
	%\email[Contact author.\\]{zoulp5@mail.sysu.edu.cn}
	\affiliation{Sino-French Institute of Nuclear Engineering and Technology, Sun Yat-Sen University, Zhuhai 519082, China} 
	
\begin{abstract}
While spin polarization from synchrotron radiation is well established, the polarization of orbital angular momentum (OAM) in such radiative processes remains elusive. We study radiation and polarization of relativistic electrons in a uniform magnetic field, focusing on OAM polarization radiation for vortex electrons which carry intrinsic OAM. The results illustrate that transition rates are asymmetric in the low-photon-energy regime, favoring OAM decrease, analogous to the spin-flip asymmetry in the Sokolov-Ternov effect. Under these conditions, synchrotron radiation can polarize the OAM. The characteristic relaxation time and stationary-state OAM distribution are obtained analytically. The polarization of spin about \(\mathcal{P}_{\text{spin}}\) reaches \(92.38\%\), while that of  \(\mathcal{P}_{\text{OAM}}\)  can even approach almost unity for a large OAM; however, their polarization behaviors are different. For typical storage ring parameters, the OAM polarization time is orders of magnitude shorter than the spin polarization time. Thus, synchrotron radiation offers a mechanism for controlling vortex electron beams which carry OAM for high-energy accelerator applications.
\end{abstract}

	
	\maketitle
	
	\section{Introduction}
	\label{sec:introduction}
	
	Synchrotron radiation, the electromagnetic emission by relativistic charged particles undergoing centripetal acceleration in magnetic fields, constitutes a cornerstone of modern high-energy physics and accelerator science. The fully relativistic quantum theory was developed by Sokolov, Ternov, and others \cite{Schwinger1949,Barkov1985,SokolovTernov1964}, who predicted the remarkable effect of spontaneous spin polarization, whereby an electron beam circulating in a storage ring can become polarized up to \(\sim 92.38\%\) with spins antiparallel to the magnetic field. This Sokolov–Ternov effect, which has been extensively studied \cite{Silenko2015,Buzzegoli2025}, arises from the spin‑dependent part of the synchrotron radiation rate and has become an essential tool for producing polarized beams in experiments \cite{Mane2005,Karimi2012}.
	
    It should be noted that the Sokolov–Ternov spin polarization limit of 
    $92.38\%$ is obtained under the assumption of uniform and constant magnetic fields, where realistic accelerator effects, such as field inhomogeneities, orbital oscillations, and quantum fluctuations associated with discrete photon emission, are not explicitly taken into account.

    The original formulation solved the Dirac equation for an electron in a uniform vertical magnetic field, obtaining the asymptotic polarization $8/(5\sqrt{3})\approx 92.38\%$, independent of energy~\cite{SokolovTernov1986}.
    In practical storage rings, however, the magnetic fields are inhomogeneous and particles traverse a distribution of orbits, which results in the depolarizing spin resonances that do not appear in the simple uniform-field model~\cite{Rossmanith1987}.
    If the magnetic environment is artificially held constant but the electron energy is large, the achievable spin polarization $\mathcal{P}_{\text{spin}}$ is generally less than $92.38\%$, and this discrepancy can be substantial due to the onset of such resonances~\cite{Mane2025}.
    Furthermore, when the electron energy is not held constant but instead decreases because of radiative losses, the final polarization also falls below the theoretical maximum, a phenomenon first quantitatively addressed by Derbenev and Kondratenko, whose formalism incorporates the spin-diffusion processes arising from quantum fluctuations of the radiation~\cite{Derbenev1973}.
    In high-energy storage rings, depolarizing resonances (particularly those of a nonlinear nature) are the dominant factors limiting the actual polarization degree, driving it far below $92.38\%$.
    Meanwhile, it is worth noting that the use of Siberian snakes in high-energy storage rings provides a nontrivial spin precession which enables longitudinal polarization at interaction points while substantially suppressing depolarization~\cite{Montague1984}.
	
    Following the idealized framework established by Sokolov and Ternov, which captures the fundamental spin dynamics in a uniform magnetic field, the present work pursues a similarly focused investigation of the underlying quantum-electrodynamic processes for orbital angular momentum (OAM). The additional complexities encountered in practical storage rings, such as depolarizing resonances and energy fluctuations, are not within the scope of this study.
    
    Vortex beams, characterized by helical phase fronts and a well-defined OAM, represent a general class of structured wavefields applicable to both light and matter waves. Optical vortex beams have been extensively investigated, leading to the development of structured light and a variety of applications in areas such as optical manipulation and quantum information processing\cite{Maruyama2022,Maruyama2023,Allen1992,Padgett2015,Lloyd2012}. The extension of the OAM concept to matter waves has enabled the realization of vortex electron beams carrying quantized OAM \cite{Bliokh2007,Zou2021,Uchida2010,Verbeeck2010,Karlovets2026,Buzzegoli2023,Zou-2025,Zou2025,Bliokh2017}. In contrast to photons, electrons are charged massive particles, and their OAM is directly coupled to external electromagnetic fields. This feature gives rise to distinct magnetic and dynamical properties, and has motivated a range of studies in electron microscopy, fundamental quantum physics and high energy nuclear and particle physics\cite{McMorran2011,Juchtmans2015,Blilok2017,Igor2022}.  One essential aspect of vortex particles is the characterization of their OAM degree of freedom and the associated polarization. In particular, the evolution of orbital angular momentum in external electromagnetic fields becomes a nontrivial problem for charged particle beams~\cite{Karlovets2026-L}. This is especially relevant in accelerator for high energy vortex electron beams, where the interplay between OAM, beam dynamics, and polarization effects remains to be systematically understood.

    While spin polarization from synchrotron radiation is well understood,  the polarization of OAM as a dynamical degree of freedom in such radiative processes is still elusive. Recent theoretical works suggest that electrons in a uniform magnetic field can occupy Landau states labeled not only by the principal quantum number $n$ and spin projection $\zeta$, but also by a discrete OAM quantum number $\ell$ \cite{FloettmannKarlovets,Karlovets2021, Karlovets2022}. This raises the fundamental question of whether synchrotron radiation, just as it polarizes spin, also induces a net polarization in the OAM of an electron beam. Addressing this question requires a unified quantum-electrodynamic treatment, which simultaneously accounts for both spin and OAM degrees of freedom. To our knowledge, this task has not been fully accomplished.
	
	In this article, we present a comprehensive theoretical analysis of radiation and polarization phenomena for relativistic electrons in a uniform magnetic field, with particular emphasis on OAM effects. Starting from the Dirac wave function in a magnetic field, the explicit expressions for transition probabilities and radiation intensities involving both spin and OAM are derived. Using the Wentzel–Kramers–Brillouin (WKB) approximation for large quantum numbers, the analytical results are obtained which reveal how synchrotron radiation polarizes electron beams not only in spin but also in OAM. Our calculations show that for low‑photon-energy emission, the OAM transition rates are asymmetric, favoring transitions that reduce the OAM quantum number, thereby driving the electron beam toward a state of net OAM polarization. The formulas for the relaxation times and the polarization dynamics of spin and OAM are provided.
	
	Throughout the paper, we use Gaussian units unless otherwise noted. Quantum states before and after a radiative transition are denoted by unprimed and primed quantum numbers, respectively (e.g., $n,\ell,\zeta$ for the initial state and $n',\ell',\zeta'$ for the final state).
	
	The paper is organized as follows. In Sec.~\ref{sec:dirac and radiation}, the  Dirac wave function for an electron in a uniform magnetic field is presented, highlighting the quantum numbers associated with spin and OAM; then the general formalism for the transition amplitude and radiation probability is developed, including the treatment of polarization states. The WKB approximation for the relevant radial matrix elements is introduced here, which leads to analytical expressions for the OAM‑transition rates in Sec.~\ref{sec:analytical}. The resulting polarization dynamics are analyzed in Sec.~\ref{sec:OAM_polarization}, where the rate equations for spin and OAM distributions are solved. Finally, the implications of our findings are discussed and summarized in Sec.~\ref{sec:discussion}. Useful integrals and some details are collected in the Appendices. Our findings bridge the well‑established theory of spin polarization with the emerging physics of vortex electron beams, offering new insights into the angular‑momentum structure of synchrotron radiation and its potential for generating and controlling structured charged‑particle beams.
%%%%
\section{A Review of Quantum Radiation Theory}
\label{sec:dirac and radiation}
\subsection{Radiation of electron in magnetic field}
\label{sec:radiation}

The eigenfunctions of the Dirac equation for an electron in a uniform, static magnetic field are presented here. This solution, well-established in the literature~\cite{SokolovTernov1964}, incorporates the spin degree of freedom through a conserved polarization operator aligned with the field. The magnetic field is taken to be along the positive $z$-axis, $\mathbf{H} = (0,0,H)$, with the corresponding vector potential in the symmetric gauge $\mathbf{A} = (-yH/2, xH/2, 0)$. The Hamiltonian $\mathcal{H}$ and the commuting polarization operator $\bm{\mu}_3$ are
\begin{equation}
	\mathcal{H} = c \bm{\alpha} \cdot \mathbf{P} + \rho_3 m_0 c^2, \qquad
	\mathbf{P} = \mathbf{p} - \frac{e}{c}\mathbf{A},
	\label{eq:hamiltonian}
\end{equation}
\begin{equation}
	\bm{\mu}_3=\bm{\sigma}_3+\rho_2\frac{(\bm{\sigma}\times\mathbf{P})_3}{m_0c},\qquad
	\text{with } k_0\bm{\mu}_3\psi=K_0\zeta\psi.
	\label{eq:mu_3}
\end{equation}
Here, $e=-|e_0|$ is the electron charge, $E$ is the electron energy, $m_0$ is the electron rest mass, $c$ is the speed of light, $k_0 = m_0 c/\hbar$, $K = E/(\hbar c)$, and $K_0 = \sqrt{K^2 - k_3^2}$. The symbols $\bm{\alpha}$ and $\rho_i$ denote the Dirac matrices, and $\bm{\sigma}$ represents the four-dimensional Pauli matrices. The eigenvalue $\zeta = \pm 1$ corresponds to the electron spin being parallel or antiparallel to the magnetic field, respectively.

The wave function solution, $\Psi(\mathbf{r},t) = \psi(\mathbf{r}) e^{-i\epsilon E t/\hbar}$, is separable in cylindrical coordinates $(r, \phi, z)$, and the separated functions satisfy Eq.~\eqref{eq:hamiltonian} and Eq.~\eqref{eq:mu_3},
\begin{equation}
	\psi(\mathbf{r}) = \psi_{\ell,k_3}(r, \phi, z) \, \mathbf{f}.
	\label{eq:psi_general}
\end{equation}
The spatial part is characterized by the OAM quantum number $\ell$ (which is associated with the $z$-component OAM, a notation that is adopted here) and the longitudinal wave number $k_3$,
\begin{equation}
	\psi_{\ell,k_3}(r, \phi, z) = \frac{1}{\sqrt{{L}}} \frac{1}{\sqrt{2\pi}} \,
	e^{i k_3 z} \,
	e^{i (\ell - 1/2) \phi},
\end{equation}
where ${L}$ is a normalization length. The four-component spinor $\mathbf{f}$ is given in Eq.~\eqref{eq:spinor}, from which the radial dependence is encoded in the functions $I_{n,s}(\rho)$, defined as
\begin{equation}
	I_{n,s}(\rho) =\left(\frac{s!}{n!}\right)^{1/2}  \, e^{-\rho/2} \, \rho^{(n-s)/2} \, L_s^{n-s}(\rho),
	\label{eq:I_ns}
\end{equation}
where $\gamma = e_0H/(2c\hbar)$,  $\rho = \gamma r^2$, $L_s^{a}(\rho)$ is the generalized Laguerre function. The quantum numbers are related by
\begin{align}
	n &= \ell + s, \quad (n, s \geq 0; \quad -\infty < \ell \leq n), \label{eq:quantum_numbers} \\
	E^2 &= (m_0 c^2)^2 + (\hbar c)^2 (k_3^2 + 4\gamma n). \nonumber
\end{align}
The integer $n$ is the principal quantum number, $s$ is the radial quantum number, and $\ell$ is related to the OAM quantum number, which corresponds to the $z$-component of the electron's total angular momentum (TAM),
\begin{equation}
	J_z \, \psi = \hbar \left( \ell - \frac{1}{2} \right) \psi.
	\label{eq:Jz}
\end{equation}
where $J_z$ is the $z$-component operator of the TAM.
	
	The emission of synchrotron radiation by a relativistic electron in a uniform magnetic field is considered while the interaction with the quantized radiation field is introduced via the time-dependent potential $U$. Starting from the Hamiltonian,
	\begin{equation}
		i\hbar \frac{\partial \Psi}{\partial t} = \mathcal{H} \Psi, \quad
		\mathcal{H} = c\bm{\alpha} \cdot \mathbf{P} + \rho_3 m_0 c^2 + U,
		\label{eq:Dirac_with_interaction}
	\end{equation}
	the interaction Hamiltonian in the second-quantized form is in Eq.~\eqref{eq:U_plus}.
	
	Applying time-dependent perturbation theory for the emission of a single photon ($\mathbf{k},\lambda$)~\cite{SokolovTernov1986}, the transition amplitude from an initial electron state $\psi_b$ to a final state $\psi_a$ is in Eq.~\eqref{eq:C_a_general}, so the transition probability per unit time follows as
	\begin{equation}
		w_{b \to a} = \frac{e_0^2}{2\pi\hbar} \int \frac{d^3\mathbf{k}}{\kappa} \,
		\delta\!\left( \frac{E_b - E_a}{\hbar c} - \kappa \right) \,
		\sum_{\lambda} \left| \overline{\bm{\alpha}} \cdot \bm{\epsilon}^{(\lambda)*} \right|^2.
		\label{eq:wba_general}
	\end{equation}
		where $\overline{\bm{\alpha}}$ is the spatial matrix element,
	\begin{equation}
		\overline{\bm{\alpha}} = \int d^3\mathbf{r} \,
		\psi_a^\dagger(\mathbf{r}) \, \bm{\alpha} \, e^{-i\mathbf{k}\cdot\mathbf{r}} \,
		\psi_b(\mathbf{r}).
		\label{eq:alpha_bar_def}
	\end{equation}
		Here, $\mathbf{k}$ is the photon wavevector with magnitude $\kappa=|\mathbf{k}|$, $\bm{\epsilon}^{(\lambda)}$ ($\lambda=2,3$) are the polarization unit vectors, the axial symmetry of the magnetic field along the $z$-axis can be exploited to simplify the calculation. The photon wavevector in spherical coordinates is
		$\mathbf{k} = \kappa (\sin\theta \cos\phi',\, \sin\theta \sin\phi',\, \cos\theta),$
	where $\theta$ is the polar angle relative to the magnetic field. As the radiation intensity is independent of the azimuthal angle $\phi'$, one can therefore set $\phi' = \pi/2$ for convenience, yielding $\mathbf{k} = \kappa(0, \sin\theta, \cos\theta)$.
	
	The linear polarization basis vectors $\bm{\epsilon}^{(2)}, \bm{\epsilon}^{(3)}$ are chosen to align with this coordinate choice:
	$\bm{\epsilon}^{(2)} = \mathbf{e}_x, \quad
		\bm{\epsilon}^{(3)} = \cos\theta \,\mathbf{e}_y - \sin\theta \, \mathbf{e}_z.$ With this basis, the polarization sum in Eq.~\eqref{eq:wba_general} for linear polarization simplifies to
	\begin{equation}
		\sum_{\lambda=2,3} \left| \overline{\bm{\alpha}} \cdot \bm{\epsilon}^{(\lambda)*} \right|^2
		= \left| \overline{\alpha}_1 \right|^2
		+ \left| \overline{\alpha}_2 \cos\theta - \overline{\alpha}_3 \sin\theta \right|^2
		\equiv \Phi_2 + \Phi_3.
		\label{eq:polarization_sum_simple}
	\end{equation}
	Here, $\overline{\alpha}_i$ ($i=1,2,3$) are components of the matrix element vector $\overline{\bm{\alpha}}$.

	\label{subsec:matrix_elements_eval}
	
	The matrix elements $\overline{\alpha}_i$ are evaluated using the Dirac wavefunctions given before. The initial and final states are labeled by their quantum numbers: $b = (n, s, \ell, \zeta, k_3)$ and $a = (n', s', \ell', \zeta', k_3')$, where $\zeta, \zeta' = \pm 1$ denote the spin projection. The spatial integrals in Eq.~\eqref{eq:alpha_bar_def} separate into longitudinal ($z$), angular ($\phi$), and radial ($r$) parts.
	
	The longitudinal integral yields momentum conservation along the field axis:
	\begin{equation}
		\frac{1}{L} \int_{-L/2}^{L/2} dz \, e^{-i z (k_3' + \kappa \cos\theta - k_3)}
		= \delta_{k_3',\, k_3 - \kappa \cos\theta}.
		\label{eq:longitudinal_integral}
	\end{equation}
	The angular and radial integrals involve the wavefunction components $I_{n,s}(\rho)$ defined in Eq.~\eqref{eq:I_ns}. The matrix elements can be expressed in terms of the radial functions and the spinor coefficients of Eq.~\eqref{eq:coefficients} (see Appendix~\ref{app:integrals}, Eqs.~\eqref{eq:A1} and \eqref{eq:A2}).
	
	\subsection{Transitions in the high-electron-energy limit}
	\label{subsec:probability_high_energy}
	
	Since the present analysis focuses on high-energy electrons, the case can be specialized to the high-energy, quasi-classical regime relevant for synchrotron radiation. The situation is considered where an electron in a positive-energy state ($\epsilon = +1$) has vanishing initial longitudinal momentum ($k_3 = 0$) by an appropriate choice of the reference coordinate system. Furthermore, the conditions are used that the quantum numbers $n$ and $n'$ are large, allowing the sum over the final radial quantum number $s'$ to be performed via the sum rule:
	\begin{equation}
		\sum_{s'=0}^\infty I_{s,s'}^2(x) = 1\label{eq:s_relation}
	\end{equation}
	In this frame, $k_3' = -\kappa \cos\theta$ from Eq.~\eqref{eq:longitudinal_integral}. The transition probability per unit time, integrated over the photon energy, is given in Eq.~\eqref{eq:wba_integral_form}. For large \(n\) and \(n^{\prime}\), it is convenient to replace the sum over the discrete change \(\nu\) by an integral: \(\sum_{\nu} \to \int_{0}^{n} d\nu\). Transforming the integration variable from $\nu$ to the dimensionless parameter $y$ via Eq.~\eqref{eq:y_definition},  Eq.~\eqref{eq:kappa_xi},   Eq.~\eqref{eq:xi_definition},  the total transition probability of spin after considering OAM transitions takes the form
	\begin{equation}
		w(\zeta,\zeta') = \frac{27}{16\pi}
		\frac{c e_0^2}{\varepsilon_0^{9/2} E \, \xi_0 \, R^2}
		\int_0^\infty \frac{y \, dy}{(1 + \xi_0 y)^4}
		\int d\Omega \,
		\mathcal{F},
		\label{eq:w_final_spin}
	\end{equation}
	and the corresponding total radiation intensity is
	\begin{equation}
		W(\zeta,\zeta') = \frac{27}{16\pi}
		\frac{c e_0^2}{\varepsilon_0^{9/2} R^2}
		\int_0^\infty \frac{y^2 \, dy}{(1 + \xi_0 y)^5}
		\int d\Omega \,
		\mathcal{F}.
		\label{eq:W_final_spin}
	\end{equation}
	Here, $R = \sqrt{n/\gamma}$ is the effective orbital radius of the electron, $\varepsilon_0=k_0^2/K^2$,   $\xi_0=3\gamma/\left(K^2 \varepsilon_0^{3/2}\right)$, $\mathcal{F}=\mathcal{F}(\theta, y; \zeta, \zeta')$ and $\mathcal{G}=\mathcal{G}(\theta, y; \zeta, \zeta';s,s')$ are derived from $\Phi$ and  Eq.~\eqref{eq:s_relation} after the variable change to $y$. The formulas for OAM-specific transitions, $w(\ell,\ell')$ and $W(\ell,\ell')$, are obtained by summing  over final spin states and averaging over initial spins without the relation in Eq.~\eqref{eq:s_relation}, as we need to maintain $\ell$ and $\ell'$ explicitly
	\begin{equation}
		w(\ell,\ell') = \frac{27}{16\pi}
		\frac{c e_0^2}{\varepsilon_0^{9/2} E \, \xi_0 \, R^2}
		\int_0^\infty \frac{y \, dy}{(1 + \xi_0 y)^4}
		\int d\Omega \,
		\frac{1}{2}\sum_{\zeta,\zeta'}\mathcal{G},
		\label{eq:w_final_OAM}
	\end{equation}
	and radiation intensity 
	\begin{equation}
		W(\ell,\ell') = \frac{27}{16\pi}
		\frac{c e_0^2}{\varepsilon_0^{9/2} R^2}
		\int_0^\infty \frac{y^2 \, dy}{(1 + \xi_0 y)^5}
		\int d\Omega \,
		\frac{1}{2}\sum_{\zeta,\zeta'}\mathcal{G}.
		\label{eq:W_final_OAM}
	\end{equation}

  \subsection{WKB approximation}
	\label{sec:WKB and spin}
	
	The WKB method, which is usually used in Schrödinger equation, can now be applied to evaluate the functions $I_{n,n'}(x)$ and $I_{s,s'}(x)$ defined in Eq.~\eqref{eq:I_ns} for large quantum numbers $n, n', s, s'$. These functions satisfy a second order differential equation of the form
	\begin{equation}
		\frac{d^2}{d x^2} \left[ x^{1/2} y(x) \right] - f_{n,n'}(x) \, x^{1/2} y(x) = 0,
		\label{eq:I_diff_eq}
	\end{equation}
	where $y(x)=I_{n,n'}(x)$, and the function $f_{n,n'}(x)$ is 
	\begin{equation}
		f_{n,n'}(x) = \frac{(n-n')^2 - 1}{4x^2} - \frac{n+n'+1}{2x} + \frac{1}{4}.
		\label{eq:f_nn}
	\end{equation}
	Equation~\eqref{eq:I_diff_eq} can be viewed as a Schrödinger-type equation with effective potential $f_{n,n'}(x)$. The classical turning points $x_0$ and $x_0'$ ($x_0 < x_0'$) are the roots of $f_{n,n'}(x)=0$. For large quantum numbers, these are given by
	\begin{equation}
		x_0 \approx (\sqrt{n} - \sqrt{n'})^2, \qquad
		x_0' \approx (\sqrt{n} + \sqrt{n'})^2.
		\label{eq:turning_points}
	\end{equation}
	
\begin{figure}[t!]
		\centering
		\includegraphics[width=1\linewidth]{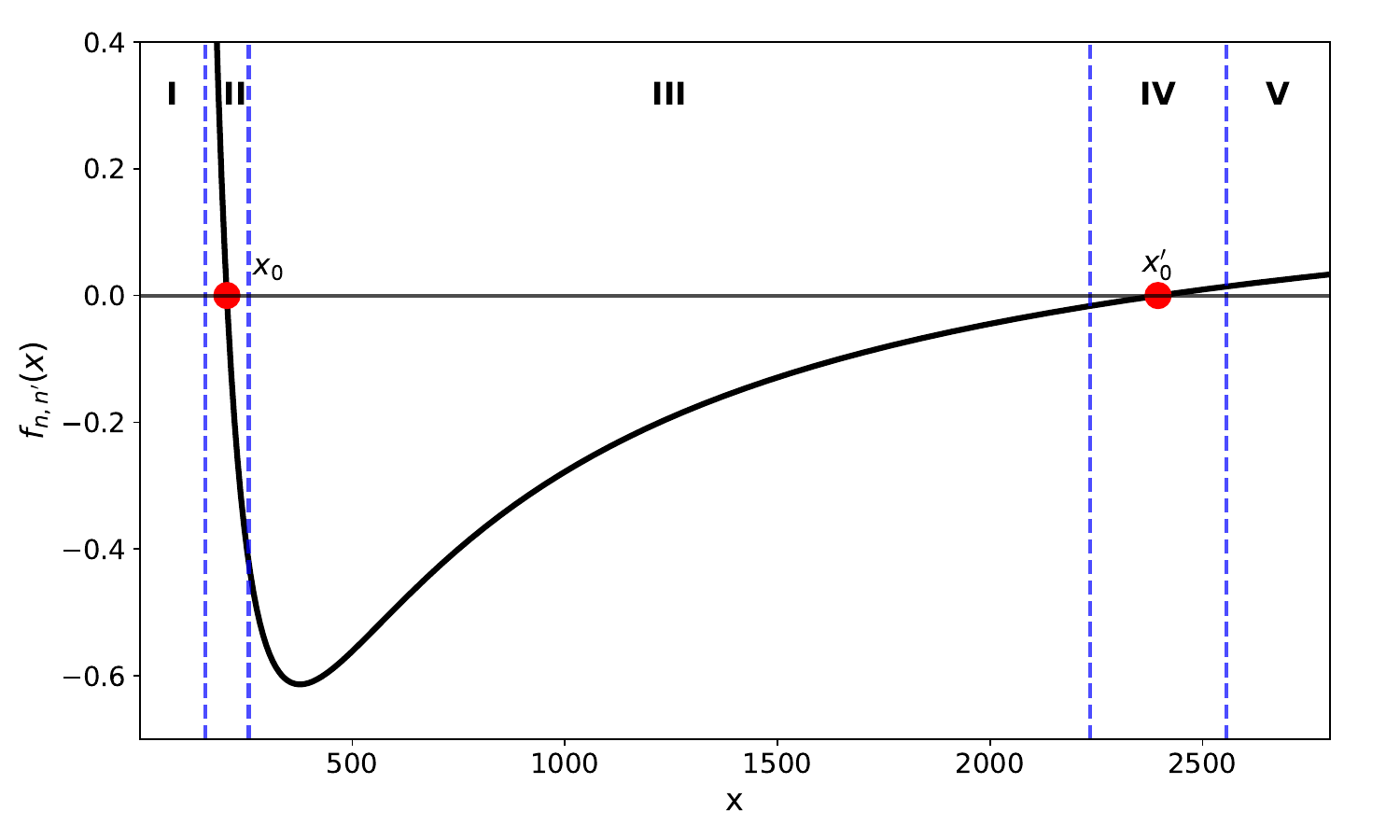}
		\caption{Effective potential $f_{n,n'}(x)$ (black solid line) from Eq.~\eqref{eq:f_nn} for $n=1000$, $n'=300$. The roots $x_0$ and $x_0'$ (red dots) are the classical turning points. The vertical dashed lines demarcate the five asymptotic regions (I–V) for the WKB solution given in Eq.~\eqref{eq:I_WKB_full}.}
		\label{fig:f_nn}
	\end{figure}

\begin{figure}[t!]
		\centering
		\includegraphics[width=1\linewidth]{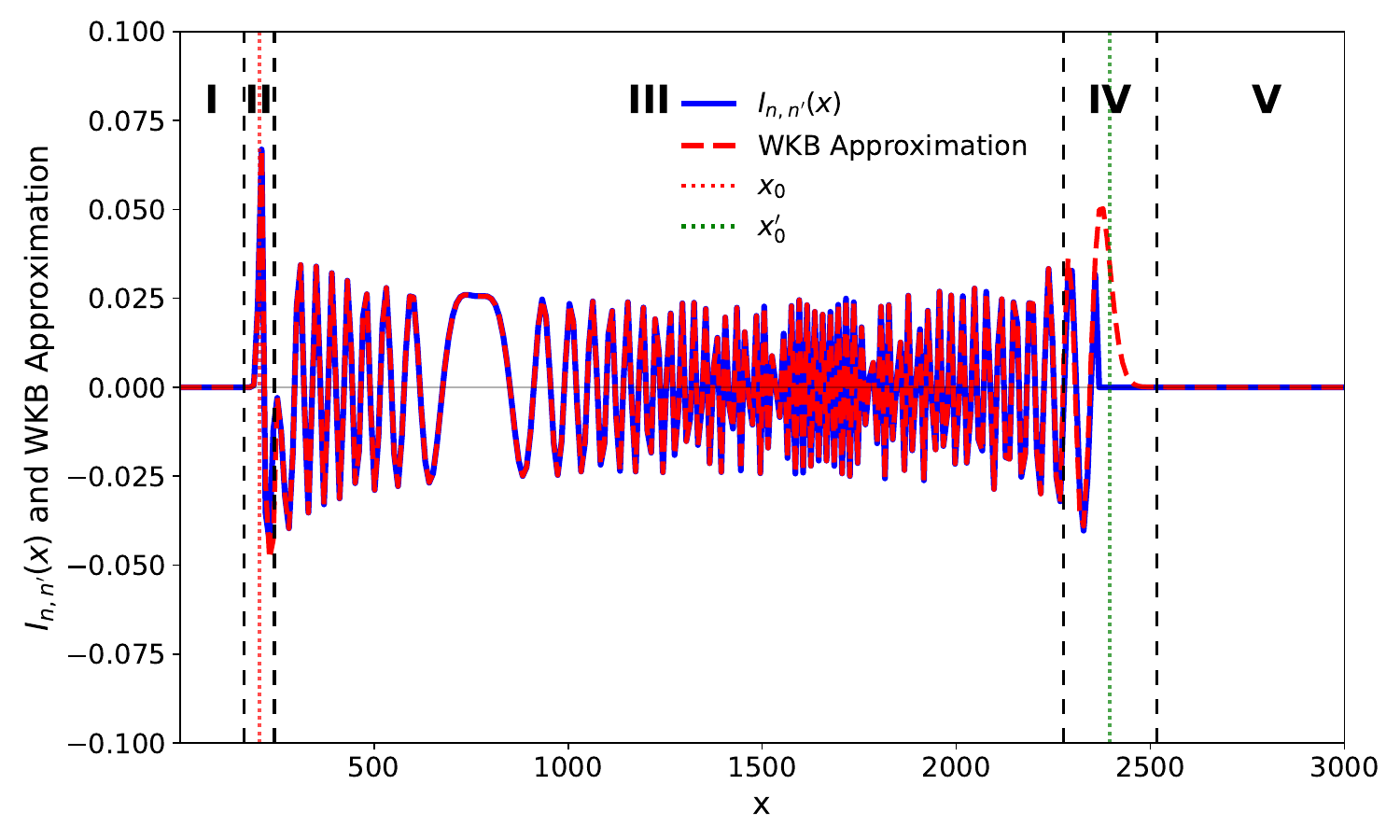}
		\caption{Comparison of the exact function $I_{n,n'}(x)$ (blue solid line) with its uniform WKB approximation (red dashed line) from  Eq.~\eqref{eq:I_WKB_full} for $n=1000$, $n'=300$. The vertical dotted lines indicate the boundaries between the asymptotic regions. The approximation accurately captures both the exponential decay (region I \& V), the transitional Airy behavior (regions II \& IV), and the oscillatory regime (region III).}
		\label{fig:I_nn_WKB}
	\end{figure}

	The argument $x = \kappa^2\sin^2\theta/(4\gamma)$ in our radiation problem, as defined in Eq.~\eqref{eq:x_definition}, is kinematically constrained. For the large quantum numbers $n,n'$, it gets  $x \approx x_0 \sin^2\theta \lesssim x_0$. Consequently, $x$ typically lies to the left of the first turning point $x_0$ and approaches it. In contrast, the functions $I_{s,s'}(x)$ involving the radial quantum numbers do not share this constraint, so the argument can span the region between and beyond both its turning points. Therefore, a uniform asymptotic approximation valid across all $x$ is required.
	%~\cite{Buzzegoli2025}
	Following the standard WKB connection procedure in Eq.~\eqref{eq:I_WKB_full}~\cite{Buzzegoli2025}, the solution $I_{n,n'}(x)$ is approximated differently in five regions defined by the turning points, as illustrated in Fig.~\ref{fig:f_nn}, while Fig.~\ref{fig:I_nn_WKB} demonstrates the agreement between the exact $I_{n,n'}(x)$ and its WKB approximation across different regions for representative quantum numbers.
	
	 The radiative polarization mechanism, i.e., Sokolov-Ternov effect in Appendix~\ref{subsec:spin_transitons}, can drive the electron beam toward a high degree of spin antiparallel to the magnetic field, independent of the initial polarization. For GeV-energy electrons in typical storage ring magnetic fields of $\sim 1$T, the polarization time \(\tau_{\text{spin}}\) in Eq.~\eqref{eq:tau_spin} is on the order of hours, which makes this effect observable in experiments with prolonged beam circulation.

	The formalism for OAM transitions mirrors the spin case but requires careful treatment of the radial quantum number sums. The key difference is that one cannot use the sum rule in Eq.~\eqref{eq:s_relation} because the final OAM quantum number $\ell'$ must be retained, which constrains the relation between $s'$ and $\ell'$ via $n' = \ell' + s'$. The WKB approximation Eq.~\eqref{eq:I_WKB_full} must therefore be applied to both $I_{n,n'}(x)$ and $I_{s,s'}(x)$ functions.

\section{Radiation and Polarization in the Low-Photon-Energy Limit}
\label{sec:analytical}

\subsection{Quasiclassical states for large quantum numbers}
\label{subsec:I_ss_approx}
To facilitate the analysis of radiative transitions in the limit of large quantum numbers, the WKB approximation is employed here to evaluate the matrix elements $I_{n,n'}(x)$ and $I_{s,s'}(x)$. The validity of this approximation requires the spacing between adjacent Landau levels to be much smaller than the electron energy, i.e., $\omega_B \ll E$~\cite{Buzzegoli2025}. For typical storage ring parameters ($E \sim 1\ \text{GeV}$, $B \sim 1\ \text{T}$), this condition is well satisfied. Moreover, since the energy of the emitted photon is directly tied to the change in the electron's Landau level index, the regime of low-photon-energy naturally emerges as the relevant one for the WKB treatment, where the quantum number change $|\Delta n|$ is small in this regime, which is consistent with the quasiclassical picture. The classical synchrotron spectrum peaks near the critical frequency $\omega_c = \frac{3}{2}\gamma^3 c/R$~\cite{Jackson1999}, but quantum effects become significant when the photon energy approaches the electron energy. In the opposite limit of low-photon-energy ($\hbar\omega \ll E$), the transition probabilities are dominated by processes with $|\Delta n|, |\Delta\ell| \sim 1$, and exhibit a systematic asymmetry that is central to the OAM polarization effect studied here. The combined considerations for the validity of WKB approximation and the spectral characteristics therefore lead us to focus on the low-photon-energy limit in the following analysis.

The regime of interest is where the radial quantum numbers \(s\) and \(s'\) are large, while the OAM quantum numbers \(\ell\) and \(\ell'\) are relatively small. Specifically, this is
\begin{equation}
	\frac{|\ell|}{n} \ll 1, \quad \frac{|\ell'|}{n'} \ll 1, \quad \frac{|\ell - \ell'|}{n} \ll 1,
	\label{eq:OAM_small_condition}
\end{equation}.

Under these conditions, the left turning point \(x_0^p\) for the function \(I_{s,s'}(x)\), defined as the root of corresponding potential \(f_{s,s'}(x)=0\), can be expressed in terms of the principal quantum number turning point \(x_0\) given in Eq.~\eqref{eq:turning_points}. Using the expansion \(\sqrt{s} = \sqrt{n - \ell} \approx \sqrt{n} \left(1 - \ell/(2n)\right)\), one obtains
\begin{align}
	x_0^p &= (\sqrt{s} - \sqrt{s'})^2 \nonumber \\
	&\approx \left[ (\sqrt{n} - \sqrt{n'}) - \frac{1}{2} \left( \frac{\ell}{\sqrt{n}} - \frac{\ell'}{\sqrt{n'}} \right) \right]^2.
	\label{eq:x0p_expansion}
\end{align}
The difference \(\sqrt{n} - \sqrt{n'}\) is related to the spectral variable \(y\) through the energy relation \(n/n' \approx (K/K')^2 = (1 + \xi_0 y)^2\) in Eq.~\eqref{eq:kappa_xi}. This yields
\begin{equation}
	\sqrt{n} - \sqrt{n'} \approx \frac{\xi_0 y \sqrt{n}}{1 + \xi_0 y}.
	\label{eq:sqrt_n_diff}
\end{equation}
Consequently, the turning point ratio can be written as $(\ell\neq \ell')$
\begin{equation}
	\frac{x_0^p}{x_0} \approx \left[ 1 + \frac{1 + \xi_0 y}{\xi_0 y} \frac{\ell' - \ell}{2n} \right]^2,
	\label{eq:x0p_ratio}
\end{equation}
where the result has neglected terms of order $\ell/n$ and used the low-photon-energy limit for the radiated photon, which implies $y$ is small.

The kinematic constraint \(x = x_0 \sin^2\theta\) places the argument of \(I_{s,s'}(x)\) near its turning point $x_0^p$ when the emitted photon energy is small. For the dominant contribution to OAM transitions, we focus on the region where \(y\) is sufficiently small such that
\begin{equation}
	y \leq \frac{|\ell - \ell'|}{2n \xi_0}.
	\label{eq:y_small_condition}
\end{equation}
In this regime, the factor \(\eta \equiv (x_0/x_0^p)\) satisfies \(\eta \ge 0\), and the quantity \((1 - x/x_0^p)\) can be expressed in terms of \((1 - x/x_0)\) and \(\eta\). Using the relation \((1 - x/x_0) \approx (1 + \xi_0 y) \varepsilon\) from Eq.~\eqref{eq:WKB_args}, one can find
\begin{equation}
	1 - \frac{x}{x_0^p} \approx \left(1 - \frac{x}{x_0}\right) \frac{1}{\varepsilon} \bigl[ 1 + 
	\eta (\varepsilon - 1) \bigr].
	\label{eq:x_relation_approx}
\end{equation}

In the WKB region II (near the first turning point), the function \(I_{s,s'}(x)\) is approximated by the modified Bessel function \(K_{1/3}\) as in Eq.~\eqref{eq:I_approx_II}:
\begin{align}
	I_{s,s'}(x) &\approx \frac{1}{\pi\sqrt{3}} \left(1 - \frac{x}{x_0^p}\right)^{1/2} K_{1/3}(z'), \label{eq:I_ss_WKB} \\
	z' &= \frac{2}{3} \bigl( (x_0^p)^2 s s' \bigr)^{1/4} \left(1 - \frac{x}{x_0^p}\right)^{3/2}. \nonumber
\end{align}
Using the relations, the argument \(z'\) can be written in terms of the variable \(z\) from Eq.~\eqref{eq:WKB_args}:
\begin{equation}
	z' \approx \left| 1 + \frac{1 + \xi_0 y}{\xi_0 y} \frac{\ell' - \ell}{2n} \right| \,
	\bigl[ 1 + \eta (\varepsilon - 1) \bigr]^{3/2} \, z.
	\label{eq:zprime_general}
\end{equation}
For \(y\) satisfying Eq.~\eqref{eq:y_small_condition}, the first factor is dominated by the OAM difference term:
\begin{equation}
	z' \approx c_0 \left| y + \frac{1 + \xi_0 y}{\xi_0} \frac{\ell' - \ell}{2n} \right| \,
	\bigl[ 1 + \eta (\varepsilon - 1) \bigr]^{3/2},
	\label{eq:zprime_simplified}
\end{equation}
where \(c_0 = (1/2) \varepsilon_0^{-3/2}\).

To verify that this approximation lies within the WKB region II, we check the condition \(x_0^p - \delta_1 < x < x_0^p + \delta_2\), where \(\delta_1, \delta_2\) are the transitional widths defined in Eq.~\eqref{eq:I_WKB_full}. For large \(n\), this condition reduces to
\begin{equation}
	\eta^{-5/6}\left[\left(\xi_0y\right)^{-1/6}\eta^{-1/6}-\left(n\xi_0y\right)^{-1/3}\right]<\left(\xi_0y\right)^{-1/6}.
	\label{eq:region_II_condition}
\end{equation}
Combined with the upper bound from Eq.~\eqref{eq:y_small_condition}, this inequality is satisfied for typical parameters \(n \gg 1\) and \(|\ell - \ell'| \sim 1\), confirming the consistency of using the region-II approximation.

\subsection{Asymptotic behavior in the low-photon-energy limit}%Limit
\label{subsec:small_y_approximation}

\begin{figure}[t!]
	\centering
	\includegraphics[width=1\linewidth]{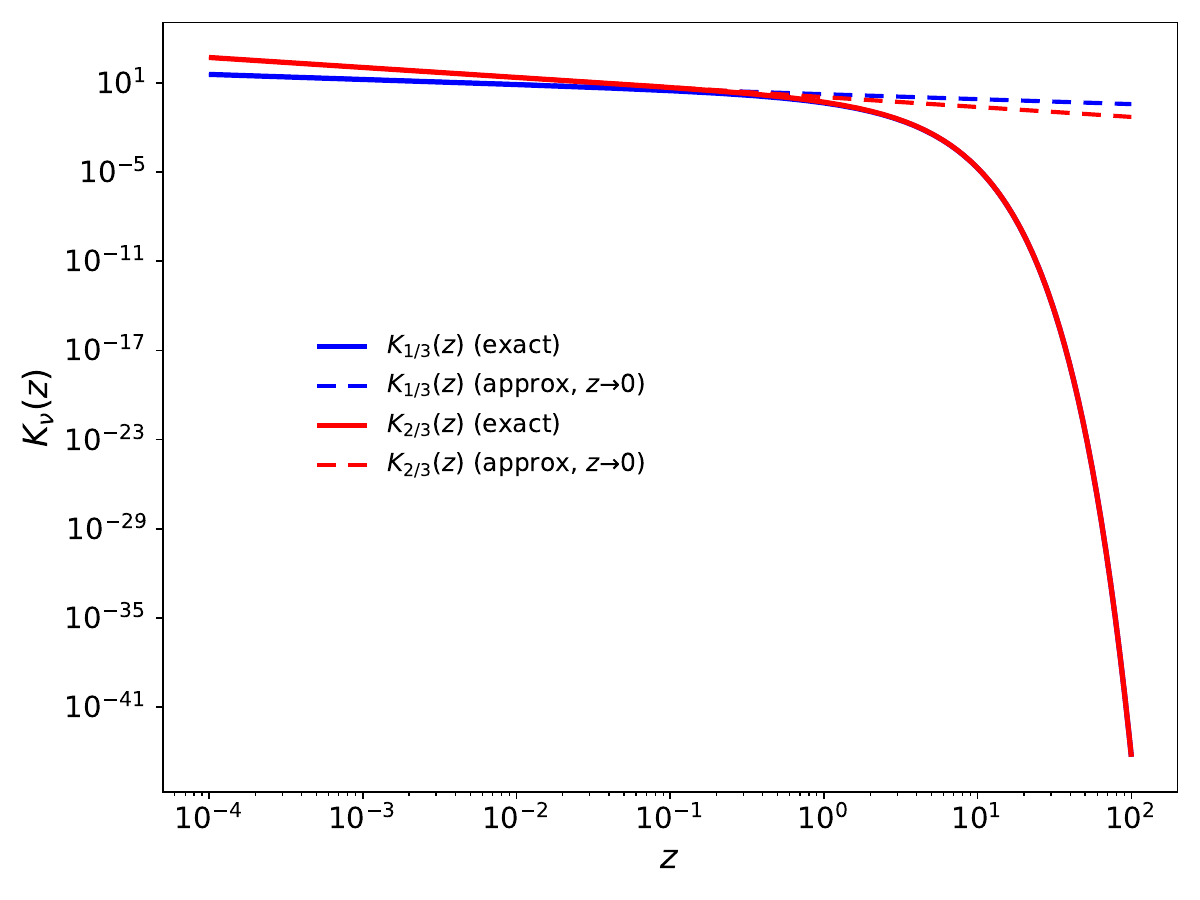}
	\caption{Modified Bessel functions \(K_{1/3}(z)\) and \(K_{2/3}(z)\) (solid lines) compared with their small-argument expansions from Eq.~\eqref{eq:K13_small_z} (dashed lines). The approximation is accurate for \(z \leq 0.1\).}
	\label{fig:K_approx}
\end{figure}

\begin{figure*}[t!]
	\centering
	\includegraphics[width=1\linewidth]{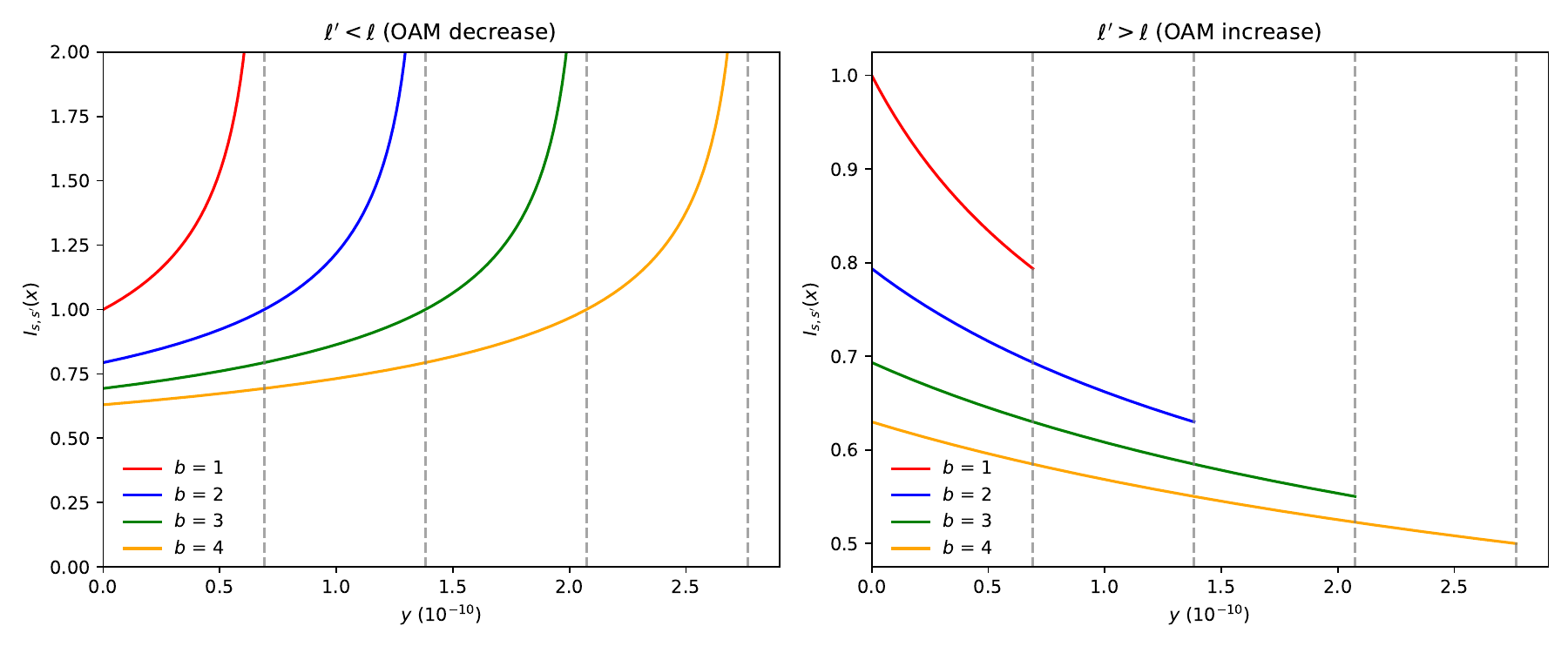}
	\caption{Comparison of the function \(I_{s,s'}(x)\) for different values of \(b = |\ell' - \ell|\). The function is normalized such that its value is unity for \(b = 1\) and \(y = 0\). The dashed lines indicate the upper bounds of \(y\) in Eq.~\eqref{eq:I_ss_final_approx} for the respective \(b\) values. The \(y\) is in units of \(10^{-10}\).}
	\label{fig:Issp-L}
\end{figure*}

The dominant contribution to OAM transitions in our case comes from the emission of soft photons, i.e., \(y \to 0\). In this limit, the modified Bessel function \(K_{\mu}(z')\) can be approximated by its small-argument expansion,
\begin{equation}
	K_{\mu}(z') \approx \frac{\Gamma(\mu)}{2} \left( \frac{2}{z'} \right)^{\mu},
	\label{eq:K13_small_z}
\end{equation}
as illustrated in Fig.~\ref{fig:K_approx}. Here,  $\Gamma(\mu)$ is the Gamma function. Substituting Eq.~\eqref{eq:zprime_simplified} into this expansion, we obtain explicit expressions for the two cases \(\ell' < \ell\) (OAM decrease) and \(\ell' > \ell\) (OAM increase):
\begin{align}
	I_{s,s'}(x) &\approx \frac{\Gamma(1/3)}{2\pi\sqrt{3}} \,
	\begin{cases}
		\displaystyle \left( \frac{2}{-c_0 y + \frac{\ell - \ell'}{3}} \right)^{1/3}, & \ell' < \ell, \\[10pt]
		\displaystyle \left( \frac{2}{c_0 y + \frac{\ell' - \ell}{3}} \right)^{1/3}, & \ell' > \ell.
	\end{cases}
	\label{eq:I_ss_final_approx}
\end{align}
The asymmetry between the two cases stems from the sign of OAM difference in the denominator of the argument. As shown in Fig.~\ref{fig:Issp-L}, for the dominant contribution corresponding to the smallest nonzero change \(|\ell - \ell^{\prime}| = 1\), Eq.~\eqref{eq:I_ss_final_approx} simplifies further to
\begin{equation}
	I_{s,s'}(x) \approx \frac{\Gamma(1/3)}{2\pi\sqrt{3}} \,
	\left( \frac{2}{|\mp c_0 y + 1/3|} \right)^{1/3},
	\label{eq:I_ss_delta_ell1}
\end{equation}
where the upper (lower) sign corresponds to \(\ell' = \ell - 1\) (\(\ell' = \ell + 1\)). This expression captures the essential singularity as \(y \to 0\), which governs the behavior of the OAM transition rates.

As the approximate form of \(I_{s,s'}(x)\) is established, one is now equipped to evaluate the transition probabilities and radiation intensities for OAM changes.

In the limit of low photon energy ($y \to 0$) and high electron energy ($\varepsilon_0 \ll 1$), the spinor coefficients from Eq.~\eqref{eq:AB_Taylor} simplify to Eq.~\eqref{eq:AB_small_y}.
Substituting these into the matrix element expressions Eqs.~\eqref{eq:alpha_12_elements} and \eqref{eq:alpha_3_element} yields our results.

The modified Bessel functions in the WKB approximation for the principal quantum number functions can be expanded for small $z$ (i.e., small $y$) using
\begin{align}
	K_{1/3}(z) &\approx \frac{\Gamma(1/3)}{2} \left( \frac{2}{z} \right)^{1/3}
	= \mathcal{A}_1 \, y^{-1/3} \left( \frac{\varepsilon_0}{\varepsilon} \right)^{1/2}, \label{eq:K13_expansion} \\
	K_{2/3}(z) &\approx \frac{\Gamma(2/3)}{2} \left( \frac{2}{z} \right)^{2/3}
	= \mathcal{A}_2 \, y^{-2/3} \left( \frac{\varepsilon_0}{\varepsilon} \right), \label{eq:K23_expansion}
\end{align}
where $\mathcal{A}_1 = \Gamma(1/3) 4^{1/3}/2$ and $\mathcal{A}_2 = \Gamma(2/3) 4^{2/3}/2$. The argument $z$ is given by Eq.~\eqref{eq:WKB_args}. In the small-$y$ limit, $K_{2/3}(z) \gg K_{1/3}(z)$, so the combinations of $I$-functions in Eq.~\eqref{eq:I_K_relation} lead to $K_{2/3}(z)$ dominating.

The transition rate for OAM changes requires the spin-averaged angular integral of $\Phi = \Phi_1 + \Phi_2$, where $\Phi_1 = |\overline{\alpha}_1|^2$ and $\Phi_2 = |\overline{\alpha}_2 \cos\theta - \overline{\alpha}_3 \sin\theta|^2$. After substituting the small-$y$ approximations and performing the angular integration over $\theta$ and $\phi$, the dominant contribution arises from $\Phi_1$. The spin-averaged result is
\begin{equation}
	\frac{1}{2} \sum_{\zeta,\zeta'} \int d\Omega \, \Phi_1
	\approx \frac{4}{3\pi} \, \mathcal{A}_2^2 \, \varepsilon_0^2 \, y^{-4/3} \, I_{s,s'}^2(x),
	\label{eq:Phi1_avg_integrated}
\end{equation}
where $d\Omega = \sin\theta d\theta d\phi$. The function $I_{s,s'}^2(x)$ is evaluated using the small-argument approximation from Eq.~\eqref{eq:I_ss_delta_ell1} for $|\ell' - \ell| = 1$ and is finally multiplied by the correction factor $1/3$ because the variation is near $1/3$ which is not following the small variation condition strictly. This gives
\begin{equation}
	I_{s,s'}^2(x) \approx \frac{1}{3\pi^2} \left( \frac{1}{2} \right)^{2/3} \mathcal{A}_1^2 \,
	\left( \frac{1}{|\mp c_0 y + 1/3|} \right)^{2/3}\frac{1}{3},
	\label{eq:Iss_sq_approx}
\end{equation}

Combining Eqs.~\eqref{eq:Phi1_avg_integrated} and \eqref{eq:Iss_sq_approx}, we define the spectral function $F(y)$ that appears in the integrand of the transition rate formula Eq.~\eqref{eq:w_final_OAM}:
\begin{equation}
	F(y) \equiv \frac{4}{27\pi^3} \left( \frac{1}{2} \right)^{2/3} \mathcal{A}_1^2 \mathcal{A}_2^2 \,
	\varepsilon_0^2 \, y^{-1/3} \,
	\left( \frac{1}{|c_0 y \pm 1/3|} \right)^{2/3}.
	\label{eq:Fy_definition}
\end{equation}
Fig.~\ref{fig:Fy_approx} compares this analytical approximation with the exact numerical evaluation of $F(y)$, confirming its accuracy in the small-$y$ region.

\subsection{Transition rates and relaxation times}
\label{subsec:transition_rates_results}

The transition rate for an OAM change $\Delta\ell = \ell' - \ell = \pm 1$, which gives the dominant contribution for the function, is obtained by integrating $F(y)$ over $y$ up to the cutoff $y_0 = |\Delta\ell|/(2n\xi_0)$ from Eq.~\eqref{eq:y_small_condition}:
\begin{equation}
	w_{\Delta\ell = \pm 1} = Q \int_0^{y_0} dy \, F(y),
	\label{eq:w_integral}
\end{equation}
where $Q = \frac{27}{16\pi} \frac{c e_0^2}{\varepsilon_0^{9/2} E \xi_0 R^2}$ is the prefactor from Eq.~\eqref{eq:w_final_OAM}. Performing the integral analytically yields
\begin{align}
	w_{+} &\equiv w_{\Delta\ell = +1} = Q \, \frac{4}{27\pi^3} \mathcal{A}_1^2 \mathcal{A}_2^2 \, \varepsilon_0^3 \, \beta_{+}, \label{eq:w_plus} \\
	w_{-} &\equiv w_{\Delta\ell = -1} = Q \, \frac{4}{27\pi^3} \mathcal{A}_1^2 \mathcal{A}_2^2 \, \varepsilon_0^3 \, \beta_{-}, \label{eq:w_minus}
\end{align}
with the integration constants $\beta_{+} \approx 1.2310$ and $\beta_{-} \approx 3.6276$. The ratio of the rates is
\begin{equation}
	\frac{w_{+}}{w_{-}} \approx 0.3393,
	\label{eq:w_ratio}
\end{equation}
which indicates that transitions with decreasing OAM ($\Delta\ell = -1$) are approximately three times more probable than those that increase it ($\Delta\ell = +1$). Similar to Eq.~\eqref{eq:w_spin_flip}, this asymmetry drives the OAM polarization of the electron beam.

For a numerical estimate, we consider typical synchrotron radiation parameters such that  electron energy $E \approx 1\,\text{GeV}$, magnetic field $B = 1\,\text{T}$ (corresponding to $H = 10^4\,\text{Gauss}$), and principal quantum number $n \approx 10^{16}$. The spin-conserving transition rate (no spin flip) is of order $w_{\text{no-flip}} \sim 10^9\,\text{s}^{-1}$, while the spin-flip rates are much smaller, $w_{\text{flip}} \sim 10^{-5}\,\text{s}^{-1}$ to $10^{-4}\,\text{s}^{-1}$ in Appendix.~\ref{subsec:spin_polarization}. The OAM transition rates in our case when emitted photon energy is small for $|\Delta\ell|=1$ are intermediate, with $w_{+} \approx 1.67 \times 10^4\,\text{s}^{-1}$ and $w_{-} \approx 4.92 \times 10^4\,\text{s}^{-1}$.

\begin{figure*}[t!]
	\centering
	\includegraphics[width=1\linewidth]{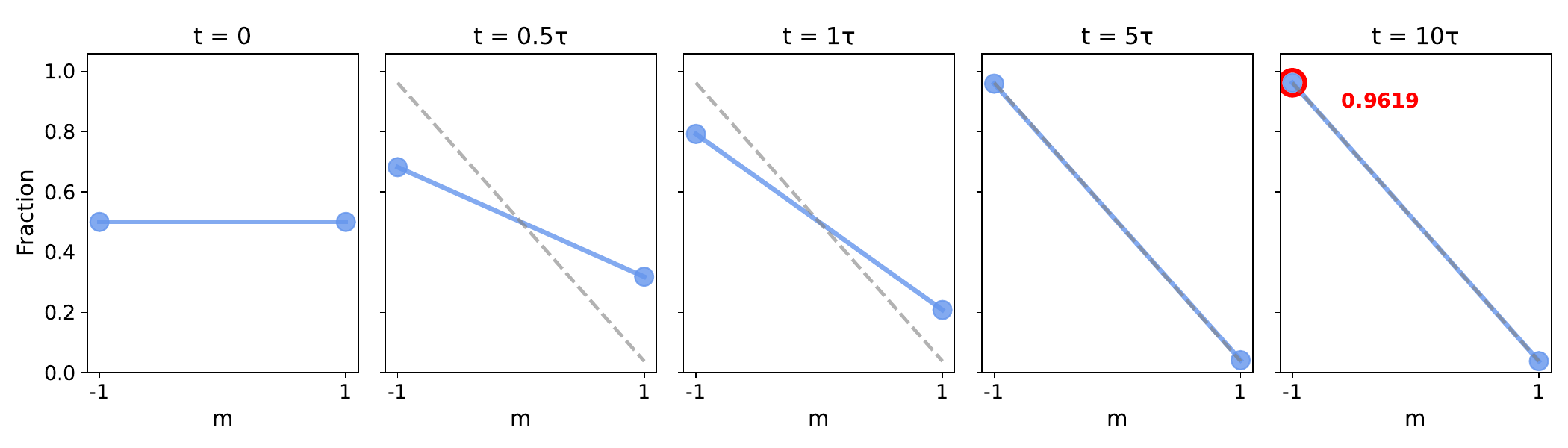}
	\caption{Time evolution of the spin population fractions \(N_{1}(t)\) (spin parallel to the magnetic field) and \(N_{-1}(t)\) (spin antiparallel), starting from an unpolarized beam. The dashed line indicates the final distribution. The system reaches an equilibrium polarization of approximately 92.38\%, corresponding to a stationary-state population of 96.19\%.}.
	\label{fig:spin_dist}
\end{figure*}

The relaxation dynamics of the spin and OAM distributions can be described by rate equations. For spin, the polarization evolves toward an equilibrium value with a characteristic time  $\tau_{\text{spin}}$ given by Eq.~\eqref{eq:tau_spin}. Fig.~\ref{fig:spin_dist} shows the time evolution of the spin populations, where one can see a stationary-state population of 96.19\% for \(N_{-1}\), which is antiparallel to the magnetic field. For OAM, the evolution depends on the initial \(\ell_0\) value, as will be detailed later. 
 
\begin{figure*}[t!]
	\centering
	\includegraphics[width=1\linewidth]{fig5.pdf}
	\caption{Comparison of the approximate spectral function \(F(y)\) (approximation) and its integral with the exact counterparts for the case \(\ell' = \ell+1\) (\(\Delta\ell = +1\)), with \(n = 10^{16}\), \(H = 10^4\) Gauss. The approximation is accurate for \(y \leq 10^{-8}\), which suffices for \(y_0 \approx 10^{-10}\).}
	\label{fig:Fy_approx}
\end{figure*}

\begin{figure*}[t!]
	\centering
	\begin{subfigure}{1\linewidth}
		\includegraphics[width=\linewidth]{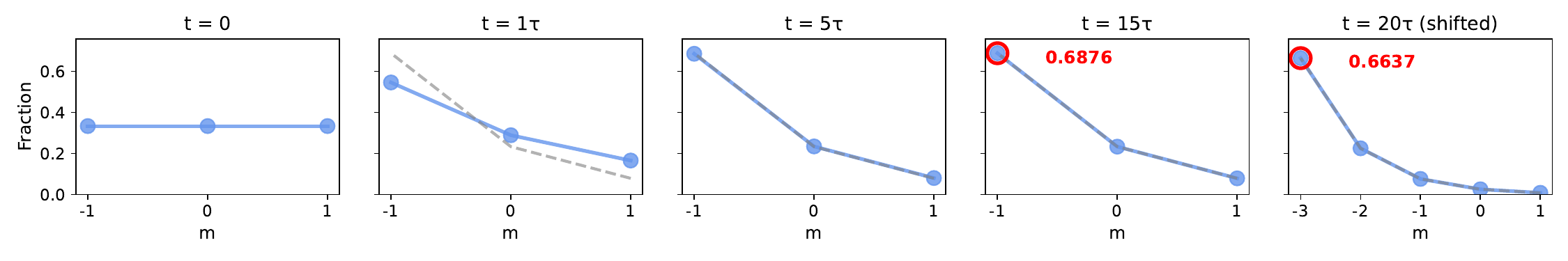}
		\caption{$\ell_{0} = 1$}
	\end{subfigure}
	\hfill
	\begin{subfigure}{1\linewidth}
		\includegraphics[width=\linewidth]{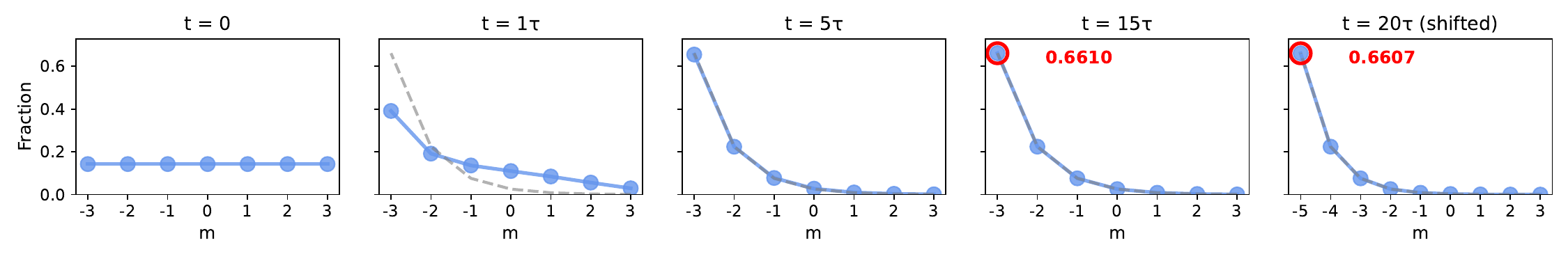}
		\caption{$\ell_{0} = 3$}
	\end{subfigure}
	\\
	\begin{subfigure}{1\linewidth}
		\includegraphics[width=\linewidth]{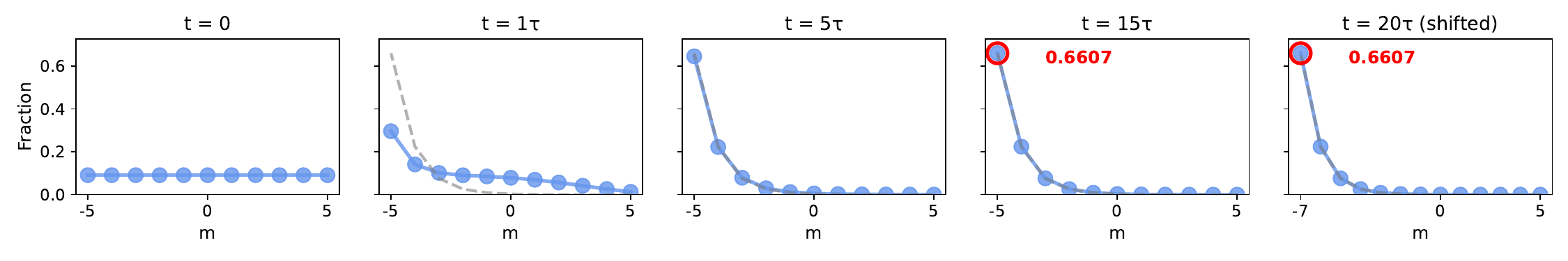}
		\caption{$\ell_{0} = 5$}
	\end{subfigure}
	\hfill
	\begin{subfigure}{1\linewidth}
		\includegraphics[width=\linewidth]{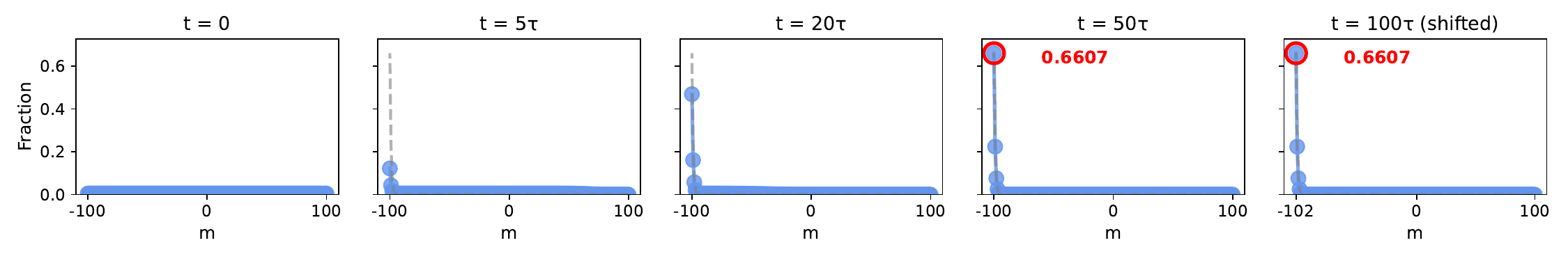}
		\caption{$\ell_{0} = 100$}
	\end{subfigure}

	\caption{Time evolution of the OAM distribution fraction for different initial \(\ell_{0}\). The plots show the population fractions of states with \(\ell = m\) for \(m = -\ell_{0}, -\ell_{0}+1, \dots, \ell_{0}-1, \ell_{0}\); the dashed line indicates the final distribution. The asymmetry between the transition rates \(w_{+}\) and \(w_{-}\) leads to a net polarization toward lower \(\ell\) values. The fifth subplot in each set (a, b, c, d) shows the final distribution when the initial particle distribution is unchanged but the lowest accessible state is \((-\ell_{0}-2)\).}
	\label{fig:OAM_dist}
\end{figure*}

\begin{figure}[t!]
	\centering
	\includegraphics[width=1\linewidth]{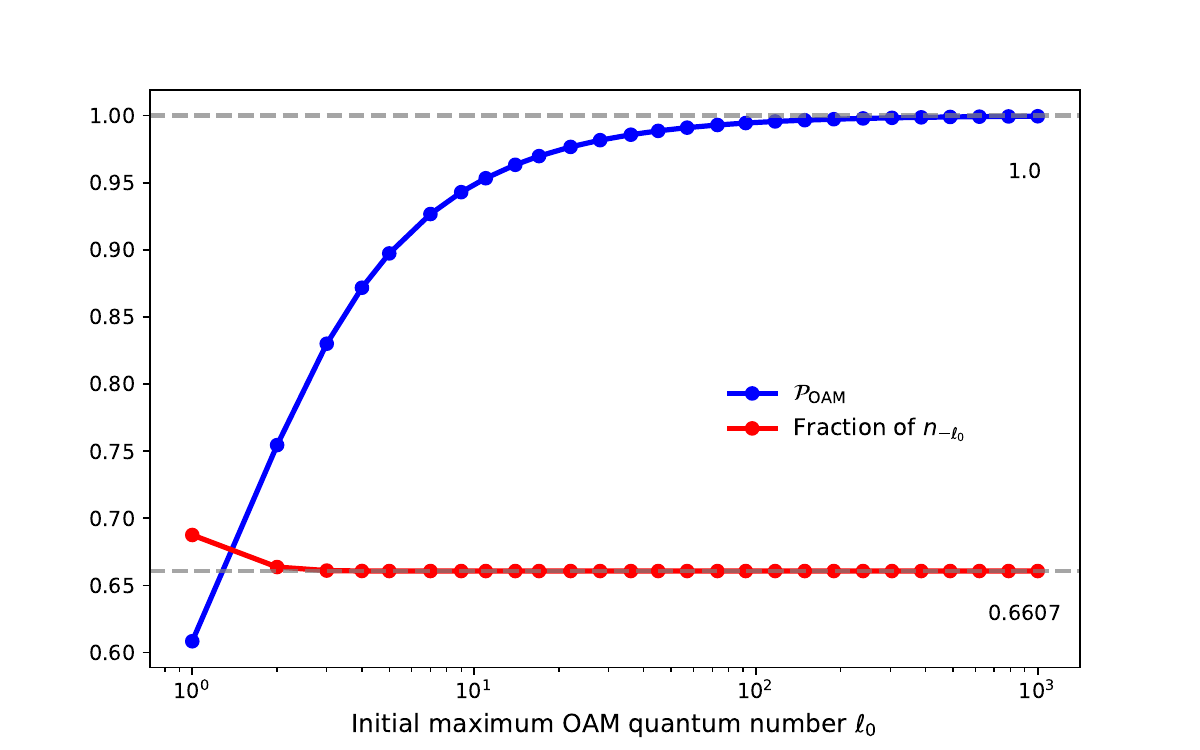}
	\caption{	\(\mathcal{P}_{\text{OAM}}\) and the fraction of the minimal $m$ state \(n_{-\ell_0}\) vary with \(\ell_0\). For large \(\ell_0\), their values approach a constant.}
	\label{fig:polarization}
\end{figure}

\begin{figure}[t!]
	\centering
	\includegraphics[width=1\linewidth]{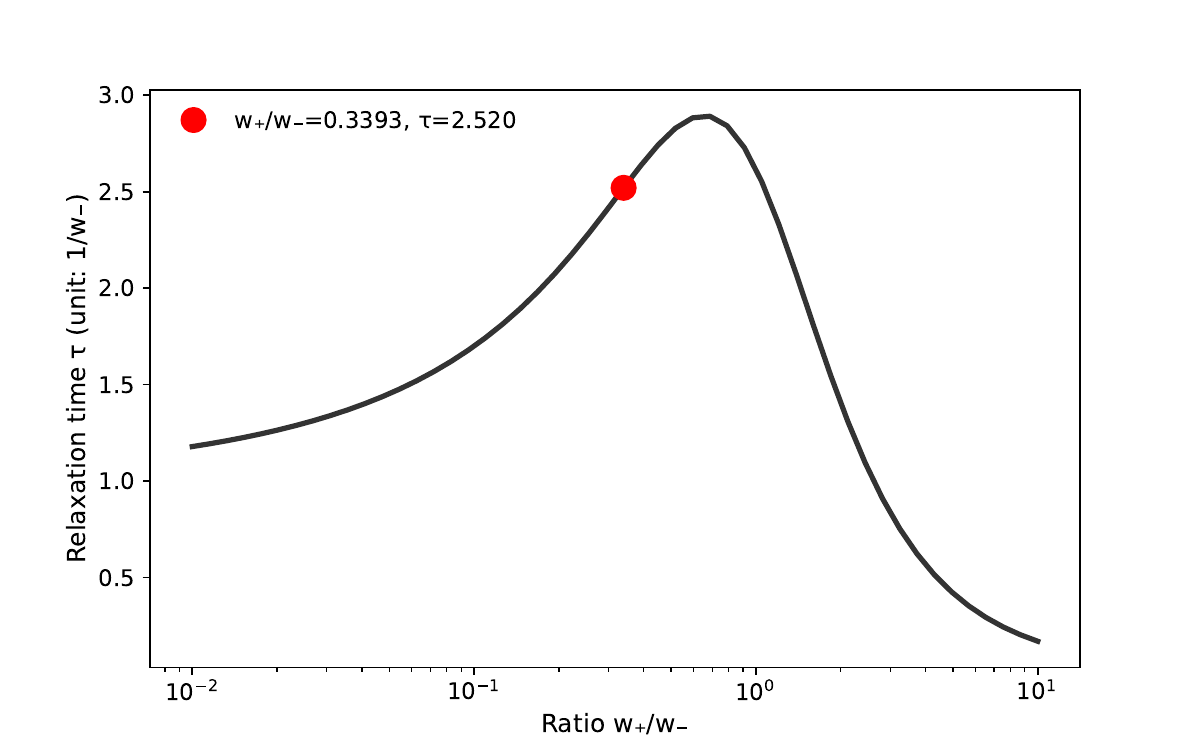}
	\caption{Variation of the OAM relaxation time $\tau_{\text{OAM}}$ with the transition rate ratio $w_{+}/w_{-}$ for an initial state with $\ell_0=2$. The relaxation time increases first and then decreases as the ratio varies.}
	\label{fig:tau_vs_ratio}
\end{figure}

The analytical results presented here provide a quantitative description of how synchrotron radiation polarizes relativistic electron beams in OAM at low-photon-energy limit. The derived transition rates and relaxation times form the basis for predicting the polarization dynamics in high-energy storage rings and astrophysical environments.
%%%%%%%%%%%%%
\subsection{OAM polarization dynamics}
\label{subsec:OAM_polarization_dynamics}

With the asymmetric transition rates for OAM-changing processes derived, the collective polarization dynamics of an ensemble of electrons under synchrotron radiation is now analyzed. This asymmetry, analogous to the spin-flip asymmetry in the Sokolov-Ternov effect~\cite{Barkov1985,SokolovTernov1964}, suggests that repeated photon emissions will drive the electron beam toward an equilibrium stationary state with a net OAM polarization. However, this polarization behavior is very different from that of spin.

To describe this evolution quantitatively, one must first specify the space of OAM states accessible to the system. In contrast to atoms, where the TAM quantum number $l$ imposes a rigid bound $|m|\le l$ on the projection $m$, the Landau states in a uniform magnetic field allow, in principle, arbitrary integer values of the OAM quantum number $\ell$~\cite{Landau Lifshitz1975}. This reflects the fact that TAM and OAM are not conserved and that $\ell$ encodes both the angular momentum projection and the guiding center position~\cite{Katoh2017}. Consequently, the OAM space is formally infinite dimensional.

In practice, however, two considerations justify truncating this space to a finite range. First, as established in Sec.~\ref{subsec:small_y_approximation}, radiative transitions dominantly couple only states with \(|\Delta\ell| = 1\). Thus, for a given initial distribution, the accessible OAM states are only those reachable via repeated \(\pm1\) steps. Second, for large \(|\ell|\), the corresponding classical orbits have guiding centers far from the axis~\cite{McMorran2011}. Such states either have negligible overlap with the initial beam profile or are kinematically suppressed in the low-photon-energy regime where our analytical rates apply (see Eqs.~\eqref{eq:OAM_small_condition} and \eqref{eq:y_small_condition}).

To better describe this polarization behavior, without loss of generality, one can therefore consider an ensemble of electrons whose OAM distribution is initially confined to a finite range \([-\ell_0, \ell_0]\), where \(\ell_0\) is the maximum allowed OAM projection in the beam, and define the relevant polarization parameters. This range is preserved under the dynamics because transitions only occur between adjacent $\ell$ states. Let $n_m(t)$ denote the population of electrons in states with OAM quantum number $\ell=m$, where $m$ ranges from $-\ell_0$ to $+\ell_0$ in integer steps. The time evolution of these populations is governed by a set of coupled rate equations that account for the incoming and outgoing fluxes for each OAM state. For interior states ($-\ell_0<m<\ell_0$), the population changes due to incoming transitions from $m+1$ and $m-1$, and outgoing transitions from $m$ to these states. For the boundary states $m=\ell_0$ and $m=-\ell_0$, only one neighboring state exists. This physical picture leads to the following system of equations~\cite{Zou2018}:

\begin{align}
	\frac{dn_{\ell_0}}{dt} &=n_{\ell_0-1} w_+ -n_{\ell_0} w_{-}, \nonumber \\
	\frac{dn_m}{dt} &= n_{m+1} w_{-} + n_{m-1} w_{+} - n_m (w_{+} + w_{-}), \nonumber\\ &(-\ell_0 < m < \ell_0)\nonumber  \\
	\frac{dn_{-\ell_0}}{dt} &= n_{-\ell_0+1} w_{-} - n_{-\ell_0} w_{+}, \label{eq:OAM_rate_eq}
\end{align}
together with the conservation of total particle number,

\begin{equation}
	\sum_{m=-\ell_0}^{\ell_0} n_m(t) = n = \text{constant}.
	\label{eq:OAM_conservation}
\end{equation}

The initial condition for an unpolarized OAM distribution is

\begin{equation}
	n_m(0) = \frac{n}{2\ell_0 + 1}, \quad m = -\ell_0, -\ell_0+1, \dots, \ell_0.
	\label{eq:OAM_initial}
\end{equation}

Equations~\eqref{eq:OAM_rate_eq}--~\eqref{eq:OAM_initial} constitute a master equation for the OAM distribution, with $w_+$ and $w_-$ serving as the transition rates per electron for increasing and decreasing the OAM projection, respectively. The truncation at \(\pm\ell_0\) is not an algebraic necessity but a requirement for better physical observation of this polarization behavior, which reflects the finite initial OAM bandwidth of the beam and the fact that transitions outside this band are suppressed by the dynamics. As will be shown, the resulting stationary distribution decays exponentially toward lower \(\ell\) and, for sufficiently large \(\ell_0\), the behavior of  polarization becomes insensitive to the precise value of \(\ell_0\).

The stationary-state solution of Eq.~\eqref{eq:OAM_rate_eq} is found to be
\begin{align}
	n_m^{\text{(st)}} &= n \, \frac{1 - \mu}{1 - \mu^{2\ell_0+1}} \, \mu^{\ell_0 - m}, \quad \mu \neq 1, \label{eq:OAM_steady_state} \\
	n_m^{\text{(st)}} &= \frac{n}{2\ell_0 + 1}, \quad \mu = 1,
\end{align}
where \(\mu \equiv w_{-}/w_{+} \approx 2.947\) from Eq.~\eqref{eq:w_ratio}. Since \(\mu > 1\), Eq.~\eqref{eq:OAM_steady_state} describes an exponential distribution favoring states with lower \(m\), i.e. the beam becomes polarized toward the minimal OAM projection state \(m = -\ell_0\). This is illustrated in Fig.~\ref{fig:OAM_dist} for various initial \(\ell_0\) values. For large \(\ell_0\), the stationary-state population of the minimal \(m\) is 66.07\%, and this value remains unchanged even when the lowest accessible state is shifted to \((-\ell_0-2)\) while the initial distribution is unchanged. This indicates that, in this case, the polarization behavior toward lower \(m\) tends to stabilize. The figure also shows that the equilibration time (in units of each \(\tau\)) increases with \(\ell_0\).

The relaxation toward the stationary state is governed by the smallest non-zero eigenvalue \(-\lambda_1\) of the transition matrix derived from Eq.~\eqref{eq:OAM_rate_eq}. The corresponding relaxation time is
\begin{equation}
	\tau_{\text{OAM}} = \frac{1}{\lambda_1},
	\label{eq:tau_OAM_def}
\end{equation}
which depends on \(\ell_0\) and the ratio \(\mu\). For GeV-energy electrons in typical storage ring magnetic fields of $\sim 1$T, the total polarization time, which is many times \(\tau_{\text{OAM}}\), is on the order of seconds to minutes and is significantly shorter than \(\tau_{\text{spin}}\). A detailed derivation of \(\lambda_1\) and its asymptotic forms is provided in Sec.~\ref{sec:OAM_polarization}.  The polarization dynamics of OAM thus exhibit a directional bias, with transitions to lower \(\ell\) states being more probable, leading to a net accumulation of electrons in states with near the minimal OAM projection. The OAM polarization is defined here analogously to the spin case, as in Eq.~\eqref{eq:spin_polarization_degree-1}, 
\begin{equation}
	\mathcal{P}_{\text{OAM}} = \frac{\sum_{m}m\cdot n_m(\infty)/n}
	{\ell_{min}}=\frac{\langle \ell \rangle}{-\ell_0}.
	\label{eq:POS_polarization_degree}	
\end{equation}  
Where $\langle \ell \rangle$ is the average $\ell$ of the electrons, $\ell_{min}$ denotes the minimum possible value of $\ell$. Fig.~\ref{fig:polarization} shows that when $\ell_0$ is large, $\mathcal{P}_{\text{OAM}}$ approaches unity. For such \(\ell_0\), the three states nearest the minimal OAM projection (TNM) account for approximately 96.09\% of the total population. In contrast to spin polarization, where the spin-flip asymmetry concentrates electrons almost entirely in a single spin state antiparallel to the field, the OAM polarization spreads over several low-\(\ell\) states. This difference arises because OAM transitions (\(|\Delta\ell| = 1\)) involve multiple accessible states and a larger rate ratio \(w_{+}/w_{-} \approx 0.3393\) (compared to \(0.0396\) for spin, Eq.~\eqref{eq:w_spin_flip}). Consequently, OAM polarization manifests as an enhanced population near the minimal projection rather than a collapse into a single quantum state. The relaxation time \(\tau_{\text{OAM}}\) for OAM polarization varies with the ratio \(w_{+}/w_{-}\), as illustrated in Fig.~\ref{fig:tau_vs_ratio} for \(\ell_0=2\). The unit of \(\tau_{\text{OAM}}\) is \(1/w_{-}\).
%%%%%%%%%%%%%%%%%%%
	\section{OAM Polarization And Relaxation TIME}
\label{sec:OAM_polarization}
%\subsection{Main equations and stationary-state solution}
\subsection{Stationary-state solution}

 One can consider the case where an electron has the maximum initial OAM projection $\ell_0$ and the range for $m = -\ell_0, -\ell_0+1, \dots, \ell_0-1, \ell_0$. Let $n_m(t)$ denote the population in state $m$, with transition rates
for $w_+$ ($m \to m+1$, increasing OAM projection) and $w_-$ ($m \to m-1$, decreasing OAM projection).

The evolution equations and the conservation of total particle number are given in Eqs.~\eqref{eq:OAM_rate_eq}-\eqref{eq:OAM_initial}. At the stationary state, all time derivatives vanish. Denoting stationary state populations by $n_m^{\text{st}}$, 
\begin{align}
	0 &= n_{\ell_0-1}^{\text{st}} w_+ - n_{\ell_0}^{\text{st}} w_- , \label{eq:OAM_steady_upper} \\
	0 &= n_{-\ell_0+1}^{\text{st}} w_- - n_{-\ell_0}^{\text{st}} w_+ , \label{eq:OAM_steady_lower} \\
	0 &= n_{m-1}^{\text{st}} w_+ + n_{m+1}^{\text{st}} w_- - n_m^{\text{st}}(w_+ + w_-)\nonumber\\
	&(-\ell_0 < m < \ell_0) . \label{eq:OAM_steady_mid}
\end{align}

Define the asymmetry parameter:
\begin{equation}
	\mu \equiv \frac{w_-}{w_+}. \label{eq:mu_def}
\end{equation}

Assume a solution of the form $n_m^{\text{st}} = C \chi^{\ell_0 - m}$ in the simplest case, where $C$ is a normalization constant and $\chi$ is a parameter to be determined. Substituting into Eq.~\eqref{eq:OAM_steady_mid} yields:
\begin{equation}
	w_+ \chi^2 - (w_+ + w_-) \chi + w_- = 0. \label{eq:chi_quadratic}
\end{equation}	
Dividing by $w_+$ and using Eq.~\eqref{eq:mu_def}:
\begin{equation}
	\chi^2 - (1 + \mu) \chi + \mu = 0. \label{eq:chi_quadratic_mu}
\end{equation}
The solutions are:
\begin{equation}
	\chi = \frac{(1+\mu) \pm |1-\mu|}{2} \quad \Rightarrow \quad \chi_1 = 1, \quad \chi_2 = \mu. \label{eq:chi_solutions}
\end{equation}
The general solution is a linear combination:
\begin{equation}
	n_m^{\text{st}} = A + B \mu^{\ell_0 - m}. \label{eq:general_solution}
\end{equation}
Applying boundary conditions \eqref{eq:OAM_steady_upper} and \eqref{eq:OAM_steady_lower} gives:
\begin{align}
	A(1 - \mu) &= 0, \label{eq:A_condition}  
\end{align}

If $\mu \neq 1$, from Eq.~\eqref{eq:A_condition}, $A = 0$. Equation \eqref{eq:A_condition} is then automatically satisfied. The solution is:
\begin{equation}
	n_m^{\text{st}} = B \mu^{\ell_0 - m}. \label{eq:solution_mu_ne1}
\end{equation}
Normalization via Eq.~\eqref{eq:OAM_conservation} determines $B$:
\begin{equation}
	B = n \frac{1 - \mu}{1 - \mu^{2\ell_0+1}}, \label{eq:B_value}
\end{equation}
yielding the stationary-state distribution:
\begin{equation}
	n_m^{\text{st}} = n \frac{1 - \mu}{1 - \mu^{2\ell_0+1}} \mu^{\ell_0 - m}, \quad \mu \neq 1. \label{eq:steady_state_mu_ne1}
\end{equation}

If $\mu = 1$ (symmetric case), equations \eqref{eq:A_condition} gives an constant. The general solution reduces to $n_m^{\text{st}} = A+B$, and normalization gives:
\begin{equation}
	n_m^{\text{st}} = \frac{n}{2\ell_0+1}, \quad \mu = 1. \label{eq:steady_state_mu_eq1}
\end{equation}

\subsection{The relaxation time from eigenvalue analysis}

The relaxation dynamics of the OAM polarization are governed by the linear system of differential equations and particle number conservation. In matrix form, these equations become:
\begin{equation}
	\frac{d\bm{n}}{dt} = M \bm{n}, \label{eq:matrix_evolution}
\end{equation}

where $\bm{n}(t) = (n_{\ell_0}, n_{\ell_0-1}, \dots, n_{-\ell_0})^\mathrm{T}$ and $M$ is the $(2\ell_0+1)\times(2\ell_0+1)$ tridiagonal matrix:
\begin{equation}
	M = 
	\begin{pmatrix}
		-w_- & w_+ & 0 & \cdots & 0 \\
		w_- & -(w_+ + w_-) & w_+ & \cdots & 0 \\
		0 & w_- & -(w_+ + w_-) & \cdots & 0 \\
		\vdots & \vdots & \vdots & \ddots & \vdots \\
		0 & 0 & \cdots & w_- & -w_+
	\end{pmatrix}. \label{eq:matrix_M_def}
\end{equation}

The formal solution of Eq.~\eqref{eq:matrix_evolution} is given by the matrix exponential:
\begin{equation}
	\bm{n}(t) = e^{Mt} \bm{n}(0). \label{eq:matrix_solution}
\end{equation}
If $M$ is diagonalizable, there exists an invertible matrix $P$ and a diagonal matrix $\Lambda = \operatorname{diag}(-\lambda_0, -\lambda_1, \dots, -\lambda_{2\ell_0})$ such that $M = P\Lambda P^{-1}$, and the solution becomes:
\begin{equation}
	\bm{n}(t) = \sum_{i=0}^{2\ell_0} c_i e^{-\lambda_i t} \bm{v}_i, \label{eq:solution_eigen_expansion}
\end{equation}
where $-\lambda_i$ are the eigenvalues of $M$, $\bm{v}_i$ are the corresponding eigenvectors, and $c_i$ are expansion coefficients determined by the initial condition.

Conservation of total particle number implies that the rows of $M$ sum to zero, i.e., $\sum_j M_{ij} = 0$ for all $i$. Consequently, $M$ has a zero eigenvalue, which we denote $-\lambda_0 = 0$, with the corresponding eigenvector proportional to the stationary-state distribution $\bm{n}^{\text{st}}$ satisfying $M\bm{n}^{\text{st}} = 0$.

All other eigenvalues have negative real parts, ensuring exponential decay of any deviation from the stationary state. Writing the solution as:
\begin{equation}
	\bm{n}(t) = \bm{n}^{\text{st}} + \sum_{i=1}^{2\ell_0} c_i e^{-\lambda_i t} \bm{v}_i, \label{eq:solution_decay}
\end{equation}

the long-time behavior is dominated by the slowest decaying mode, i.e., the eigenvalue with the largest real part among the negative eigenvalues (closest to zero). Denoting this eigenvalue as $-\lambda_1$, the relaxation time $\tau$ is defined as:
\begin{equation}
	\tau \equiv \frac{1}{\lambda_1}. \label{eq:tau_def}
\end{equation}

To determine $\lambda_1$, we solve the eigenvalue problem $M\bm{v} = -\lambda \bm{v}$. Writing the components of $\bm{v}$ as $v_m$, for $\ell_0 = 1$, the system has three OAM states $m = 1, 0, -1$.  
The rate matrix $M$ is
\begin{equation}
	M = \begin{pmatrix}
		-w_- & w_+ & 0 \\[2pt]
		w_- & -(w_+ + w_-) & w_+ \\[2pt]
		0 & w_- & -w_+
	\end{pmatrix}.
\end{equation}
The characteristic polynomial $\det(M + \lambda I)=0$ yields
\begin{equation}
	\lambda\bigl[\lambda^2 - 2(w_++w_-)\lambda + (w_+^2 + w_+w_- + w_-^2)\bigr] = 0,
\end{equation}
Thus one eigenvalue is $\lambda_0 = 0$; the other two satisfy
\begin{equation}
	\lambda^2 - 2(w_++w_-)\lambda + (w_+^2 + w_+w_- + w_-^2) = 0,
\end{equation}
which gives
\begin{equation}
	\lambda_{1,2} = (w_++w_-) \pm \sqrt{w_+w_-},
\end{equation}
The smallest positive root (which gives the slowest decay) is
\begin{equation}
	\lambda = (w_++w_-) - \sqrt{w_+w_-},
\end{equation}
Hence the OAM relaxation time is
\begin{equation}
	\tau_{\mathrm{OAM}} = \frac{1}{\lambda} = \frac{1}{(w_++w_-) - \sqrt{w_+w_-}}.
\end{equation}

In the general asymmetric case (\(w_+ \neq w_-\)), the parameters must be specified so that the eigenvalues can be computed numerically. The relaxation time is still given by Eq.~\eqref{eq:tau_def}, where \(-\lambda_1\) is the eigenvalue with the largest real part among the negative eigenvalues (closest to zero). This follows directly from the expansion \eqref{eq:solution_decay}, as the mode with eigenvalue \(-\lambda_1\) decays slowest and thus dominates the long-time approach to stationary state, which can also be proved for the spin case in Eqs.~\eqref{eq:spin_rate_eq-1-1} and \eqref{eq:spin_rate_eq-2-1}, and the same results are obtained for Eqs.~\eqref{eq:spin_solution-1} to \eqref{eq:spin_avg_solution-1}.
	%%%%%%%%%%%%%%%%%%%%%%%%%%%%%%%%%%%%%%%%%%%%%%%%%
	
	\section{Discussion and Summary}
	\label{sec:discussion}
	
	We have presented a comprehensive theoretical analysis of radiation and polarization phenomena for relativistic electrons in a uniform magnetic field, with particular emphasis on the role of OAM. Starting from the Dirac wave function, we derived expressions for transition probabilities and radiation intensities involving both spin and OAM degrees of freedom. In the high-electron-energy and low-photon-energy limits, the WKB approximation yields analytical results that reveal how synchrotron radiation polarizes electron beams in OAM.
	
    Our key findings illustrate that, in the low-photon-energy limit, synchrotron radiation induces an asymmetry in OAM transitions similar to that in spin-flip transitions, leading to spontaneous OAM polarization just as in the spin case. Unlike spin, however, the OAM quantum number in a magnetic field has in principle no lower bound, yet for physically allowed large negative \(-\ell_0\) the polarization behavior stabilizes, and the final distribution differs from that of spin. Meanwhile, the relaxation time \(\tau_{\mathrm{OAM}}\) is significantly shorter than \(\tau_{\mathrm{spin}}\). The stationary state polarization $\mathcal{P}_{\text{spin}}$ reaches about \(92.38\%\), corresponding to $N_{-1}$ of $96.19\%$, with a characteristic relaxation time given by Eq.~\eqref{eq:tau_spin}. In the case of small photon energy, the OAM transition rates between adjacent \(\ell\) states are asymmetric, with \(w_{+}/w_{-} \approx 0.3393\). This asymmetry drives the electron distribution toward states with minimal OAM projection, leading to an OAM polarization \(\mathcal{P}_{\mathrm{OAM}}\) of unity and TNM of \(96.09\%\) for large $\ell_{0}$, which indicates that the electrons ultimately become polarized near the minimal OAM projection states. The derived analytical formulas, particularly Eqs.~\eqref{eq:w_plus} and \eqref{eq:w_minus} for OAM transition rates, provide a quantitative basis for predicting polarization dynamics in high-energy electron beams.

    These results extend the well-known Sokolov-Ternov effect to the OAM degree of freedom. The polarization of spin and OAM suggests new possibilities for generating and controlling vortex electron beams with defined angular momentum properties. Such beams could find applications in advanced accelerator science and high energy physics. Our work provides the foundational radiation theory necessary to understand how such vortex states evolve and polarize under synchrotron emission. Building on the numerous existing studies on the generation of vortex electron beams~\cite{Clark2013,Zou2024,Schattschneider2012,Schattschneider2014,Mafakheri2017}, our work provides a theoretical foundation for the future production of polarized vortex electron beams. The polarization behavior under other conditions will be investigated in our future study, along with the experimental signatures of OAM-polarized beams in storage rings and the interplay between spin and OAM polarization in tailored magnetic field configurations

%%%%%%%%%%%%%%ack
    \section*{acknowledgments}
	\begin{acknowledgments}
	The work was supported by the National Key R\&D Program of China No. 2024YFE0109802, the National Natural Science Foundation of China (Grants No.~12175320 and No.~12375084), and Guangdong Basic and Applied Basic Research Foundation (Grant No.~2022A1515010280).
	\end{acknowledgments}
	
	\appendix
	\section{Useful Definitions and Integrals}
	\label{app:integrals}
	
	The four-component spinor $\mathbf{f}$ in  Eqs.~\eqref{eq:psi_general} is
	\begin{equation}
		\mathbf{f} = \left(2\gamma\right)^{1/2}
		\begin{pmatrix}
			C_1 \, I_{n-1, s}(\rho) \, e^{-i\phi/2} \\[2pt]
			i C_2 \, I_{n, s}(\rho) \, e^{i\phi/2} \\[2pt]
			C_3 \, I_{n-1, s}(\rho) \, e^{-i\phi/2} \\[2pt]
			i C_4 \, I_{n, s}(\rho) \, e^{i\phi/2}
		\end{pmatrix},
		\label{eq:spinor}
	\end{equation}
	where $\gamma = e_0H/(2c\hbar)$ and $\rho = \gamma r^2$. The coefficients $C_i$ are determined by the spin projection $\zeta$ and the kinetic parameters 
	\begin{equation}
		\begin{pmatrix}
			C_1 \\ C_2 \\ C_3 \\ C_4
		\end{pmatrix}
		= \frac{1}{2\sqrt{2}}
		\begin{pmatrix}
			B_3 (A_3 + A_4) \\
			B_4 (A_4 - A_3) \\
			B_3 (A_3 - A_4) \\
			B_4 (A_3 + A_4)
		\end{pmatrix},
		\label{eq:coefficients}
	\end{equation}
	with
	\begin{align}
		A_3 &= \left(1 + \frac{\epsilon k_3}{K}\right)^{1/2}, \quad
		A_4 = \epsilon \zeta \left(1 - \frac{\epsilon k_3}{K}\right)^{1/2}, \nonumber \\
		B_3 &=\left(1 + \frac{\zeta k_0}{K_0}\right)^{1/2} , \quad
		B_4 =\zeta \left(1 - \frac{\zeta k_0}{K_0}\right)^{1/2} .
		\label{eq:AB_coeffs}
	\end{align}
	Here, $\epsilon = \pm 1$ labels the positive or negative-energy solutions, $k_0 = m_0 c/\hbar$, $K = E/(\hbar c)$, and $K_0 = \sqrt{K^2 - k_3^2}$.
	
	And the interaction Hamiltonian in Eqs.~\eqref{eq:Dirac_with_interaction} is
		\begin{align}
		U &= U^{(+)} + U^{(-)}, \nonumber \\
		U^{(-)} &= \frac{e_0}{\mathbf{L}^{3/2}} \sum_{\mathbf{k},\lambda}
		\sqrt{\frac{2\pi \hbar c}{\kappa}} \,
		\bm{\alpha} \cdot \bm{\epsilon}^{(\lambda)} \,
		a_{\mathbf{k},\lambda} \,
		e^{-i c \kappa t + i\mathbf{k}\cdot\mathbf{r}},  \nonumber\\
		U^{(+)} &= \frac{e_0}{\mathbf{L}^{3/2}} \sum_{\mathbf{k},\lambda}
		\sqrt{\frac{2\pi \hbar c}{\kappa}} \,
		\bm{\alpha} \cdot \bm{\epsilon}^{(\lambda)*} \,
		a_{\mathbf{\mathbf{k}},\lambda}^\dagger \,
		e^{i c \kappa t - i\mathbf{k}\cdot\mathbf{r}}. \label{eq:U_plus}
	\end{align}
 Here $a_{\mathbf{k},\lambda}$, $a_{\mathbf{k},\lambda}^\dagger$ are the photon annihilation and creation operators, respectively. The quantization volume is $\mathbf{L}^3$. The transition amplitude from an initial electron state $\psi_b$ to a final state $\psi_a$ is~\cite{SokolovTernov1986}
 \begin{equation}
 	C_{a}(t) = -\frac{i e_0}{\hbar \mathbf{L}^{3/2}} \sqrt{\frac{2\pi \hbar c}{\kappa}}
 	\int_0^t dt' e^{-i (E_b - E_a - \hbar c \kappa) t'/\hbar} \,
 	\overline{\bm{\alpha}} \cdot \bm{\epsilon}^{(\lambda)*},
 	\label{eq:C_a_general}
 \end{equation}
%%%%%%%%%%
	
	For large $n, n'$, the approximation of $I$-functions follows
	\begin{equation}
		\begin{pmatrix}
			I_{nn'} \\
			I_{n,n'-1} \\
			I_{n-1,n'} \\
			I_{n-1,n'-1}
		\end{pmatrix}
		= 2N(\varepsilon)^{1/2} \left[ K_{1/3} + 
		\begin{pmatrix}
			0 \\
			-(1+\xi_0 y) \\
			1 \\
			-\xi_0 y
		\end{pmatrix}
		\varepsilon^{1/2} K_{2/3} \right]
		\label{eq:I_K_relation}
	\end{equation}
	where the normalization factor and arguments are
	\begin{align}
		N &= \frac{(1+\xi_0 y)^{1/2}}{2\pi \sqrt{3}}, \quad
		z = \frac{y}{2} \left( \frac{\varepsilon}{\varepsilon_0} \right)^{3/2},\nonumber\\
		\varepsilon &= 1 - \beta^2 \sin^2\theta, \quad
		\varepsilon_0 = 1 - \beta^2.
		\label{eq:WKB_args}
	\end{align}
	Here $\beta = v/c$ is the electron speed, $\xi_0$ is defined in Eq.~\eqref{eq:xi_definition},  and $y$ is the spectral variable from Eq.~\eqref{eq:y_definition}.
	
	The spinor coefficients $A_i$, $B_i$ in the high-electron-energy limit ($E \gg m_0 c^2$, $\varepsilon_0^{1/2} \ll 1$) are expanded to first order in $\varepsilon_0^{1/2}$
	\begin{align}
		\begin{pmatrix} A_3 \\ A_4 \end{pmatrix} &=
		\begin{pmatrix} 1 \\ \zeta \end{pmatrix}, \quad
		\begin{pmatrix} B_3 \\ B_4 \end{pmatrix} =
		\begin{pmatrix} 1 \\ \zeta \end{pmatrix} \left( 1 \pm \frac{1}{2} \zeta \varepsilon_0^{1/2} \right), \nonumber \\
		\begin{pmatrix} A_3' \\ A_4' \end{pmatrix} &=
		\begin{pmatrix} 1 \\ \zeta' \end{pmatrix} \left( 1 \mp \frac{1}{2} \xi_0 y \cos\theta \right), \nonumber \\
		\begin{pmatrix} B_3' \\ B_4' \end{pmatrix} &=
		\begin{pmatrix} 1 \\ \zeta' \end{pmatrix} \left( 1 \pm \frac{1}{2} \zeta' \varepsilon_0^{1/2} (1 + \xi_0 y) \right).
		\label{eq:AB_Taylor}
	\end{align}

%%%%%%%%%%%%%%%%%%%
In the high-electron-energy and low-photon-energy limits, one can  get from Eq.~\eqref{eq:AB_Taylor}
\begin{align}
	\begin{pmatrix} A_3 \\ A_4 \end{pmatrix} &= \begin{pmatrix} 1 \\ \zeta \end{pmatrix}, \quad
	\begin{pmatrix} B_3 \\ B_4 \end{pmatrix} = \begin{pmatrix} 1 \\ \zeta \end{pmatrix}, \nonumber \\
	\begin{pmatrix} A_3' \\ A_4' \end{pmatrix} &\approx \begin{pmatrix} 1 \\ \zeta' \end{pmatrix}
	\left( 1 \mp \frac{1}{2} \xi_0 y \cos\theta \right), \nonumber \\
	\begin{pmatrix} B_3' \\ B_4' \end{pmatrix} &\approx \begin{pmatrix} 1 \\ \zeta' \end{pmatrix}.
	\label{eq:AB_small_y}
\end{align}
the dimensionless parameter $y$ is defined as
\begin{equation}
	y = \frac{K \kappa \varepsilon_0^{3/2}}{3\gamma\left(1 - \kappa/(K-k_0)\right)},
	\quad \text{with } \varepsilon_0 = \frac{k_0^2}{K^2} = \left(\frac{m_0 c^2}{E}\right)^2,
	\label{eq:y_definition}
\end{equation}
 it simplifies the expressions. In the high-electron-energy limit ($E \gg m_0 c^2$, $\varepsilon_0^{1/2} \ll 1$), the relations can be obtained
\begin{align}
	\kappa &= \frac{K \xi_0 y}{1 + \xi_0 y}, \qquad
	K' = \frac{K}{1 + \xi_0 y}\label{eq:kappa_xi}\\
	d\kappa &= \frac{\xi_0 K}{(1 + \xi_0 y)^2} dy, \quad
	\xi_0 \equiv \frac{3\gamma}{K^2 \varepsilon_0^{3/2}}
	= \frac{3}{2} \frac{H}{H_0} \frac{E}{m_0 c^2},
	\label{eq:xi_definition}
\end{align}
where $H_0 = m_0^2 c^3/(e_0\hbar)$ is the Schwinger critical field.

	Substituting Eqs.~\eqref{eq:I_K_relation}, \eqref{eq:AB_Taylor} into the expressions for $\Phi_2$ and $\Phi_3$ [Eq.~\eqref{eq:polarization_sum_simple}], and
inserting it into the general transition rate formula Eq.~\eqref{eq:w_final_spin} while using the integral properties, one can get the result in Eqs.~\eqref{eq:w_spin_flip} and obtain
%%%%%%%%%%%%%%	
\begin{widetext}
	\begin{align}
		\begin{pmatrix} -i\overline{\alpha}_1 \\ \overline{\alpha}_2 \end{pmatrix}
		&= \frac{1}{4} \Bigl[ (A_3'A_4 + A_4'A_3)
		\bigl( B_3'B_4 \, I_{n,n'-1}(x) \mp B_4'B_3 \, I_{n-1,n'}(x) \bigr) \Bigr]
		I_{s,s'}(x), \label{eq:alpha_12_elements} \\
		\overline{\alpha}_3 &= \frac{1}{4} \Bigl[ (A_3'A_3 - A_4'A_4)
		\bigl( B_3'B_3 \, I_{n-1,n'-1}(x) + B_4'B_4 \, I_{n,n'}(x) \bigr) \Bigr]
		I_{s,s'}(x). \label{eq:alpha_3_element}
	\end{align}
\end{widetext}
The arguments $A_i, B_i$ and $A_i', B_i'$ refer to the initial and final states, respectively, and are given by Eqs.~\eqref{eq:AB_coeffs}. The variable $x$ is defined as
\begin{equation}
	x = \frac{\kappa^2 \sin^2\theta}{4\gamma},
	\label{eq:x_definition}
\end{equation}

The related transition probability per unit time is
	\begin{align}
		w(n,s,\zeta \to n',s',\zeta') &=
		\frac{e_0^2}{2\pi\hbar}\int_0^\infty d\kappa \, \kappa \,
		\delta\!\left( \kappa + K'(\kappa) - K \right)\nonumber\\
		&\times\int d\Omega \,
		\Phi(\kappa, \theta; \zeta, \zeta';s,s'),
		\label{eq:wba_integral_form}
	\end{align}
	where $K = E/(\hbar c)$, $K'(\kappa) = \sqrt{k_0^2 + 4\gamma n' + \kappa^2 \cos^2\theta}$, $d\Omega = \sin\theta \, d\theta d\phi$, and $\Phi = \Phi_2 + \Phi_3$ from Eq.~\eqref{eq:polarization_sum_simple}.
	
	The Dirac delta function enforces energy conservation, $E = E' + \hbar c \kappa$. Defining $\kappa_0$ as the positive root of $\kappa + K'(\kappa) - K = 0$, it gets
	\begin{equation}
		\kappa_0 = \frac{K - \sqrt{K^2 - 4\gamma \nu \sin^2\theta}}{\sin^2\theta},
		\label{eq:kappa0_root}
	\end{equation}
	and the derivative
	\begin{equation}
		\left. \frac{d}{d\kappa} (\kappa + K'(\kappa) - K) \right|_{\kappa=\kappa_0}
		= 1 + \frac{\kappa_0 \cos^2\theta}{K'(\kappa_0)}.
		\label{eq:delta_derivative}
	\end{equation}
	Where $\nu=n-n'$. Using the delta function identity, the $\kappa$-integral in Eq.~\eqref{eq:wba_integral_form} is performed analytically.
	
%%%%%%%%%%%%%%%%%%%%%%%%%
	The following integrals involving modified Bessel functions appear in the evaluation of the angular and spectral integrals in the main text:
	\begin{align}
		&\frac{1}{2\pi}\int_{0}^{2\pi}\mathrm{d}\phi\int_{0}^{\infty}r\mathrm{d}r exp\left\{-i\kappa r\sin\theta\sin\phi+i\left(\ell-\ell'\right)\phi\right\}\nonumber\\
		&=J_{\ell-\ell'}(\kappa r\sin\theta)
		\label{eq:A1}\\
		&\int_{0}^{\infty}J_{\ell-\ell'}\left\{2\left(x\rho\right)^{1/2}\right\}I_{n',s'}(\rho)I_{n,s}(\rho)\mathrm{d}\rho=I_{n,n'}(x)I_{s,s'}(x)
		\label{eq:A2}
	\end{align}
	Where $\rho=\gamma r^2$, $x=\left(\kappa^2\sin^2\theta\right)/\left(4\gamma\right)$, $n'=\ell'+s',n=\ell+s$, and $J_{\ell-\ell'}$ is the Bessel function. Integral formula for $y$
	\begin{align}
		&\int_{0}^{y_0}\frac{1}{y^{1/3}}\left(\frac{1}{a y+\frac{1}{3}}\right)^{2/3}dy=\frac{1}{2a^{2/3}}\ln\left(\frac{1+u_0+u_0^2}{\left(1-u_0\right)^2}\right)+\nonumber\\
		&\frac{\sqrt{3}}{a^{2/3}}\left(\frac{\pi}{6}-\arctan\left(\frac{2u_0+1}{\sqrt{3}}\right)\right),   u_0=\left(\frac{ay_0}{ay_0+1/3}\right)^{1/3}.
		\label{eqint4-b-9}
	\end{align}
	
	\section{WKB-Related Formulas}
	\label{app:WKB}
	
	The uniform asymptotic formula of WKB approximation is~\cite{Buzzegoli2025}
	\begin{widetext}
		\begin{equation}
			I_{n,n'}(x) \approx
			\begin{cases}
				\displaystyle
				\frac{1}{\sqrt{2\pi}} \, \frac{1}{(x_0 - x)^{1/4}(x_0' - x)^{1/4}} \,
				\exp\!\left[ -\mathcal{S}_1(x) \right], & \text{I: } x \le x_0 - \delta_1, \\[12pt]
				\displaystyle
				\frac{x_0^{1/3}}{(n n')^{1/12}} \, \frac{1}{x^{1/2}} \,
				\operatorname{Ai}\! \left( \frac{(n n')^{1/6}}{x_0^{2/3}} (x_0 - x) \right), & \text{II: } x_0 - \delta_1 < x < x_0 + \delta_2, \\[12pt]
				\displaystyle
				\sqrt{\frac{2}{\pi}} \, \frac{1}{(x - x_0)^{1/4}(x_0' - x)^{1/4}} \,
				\sin\!\left[ \mathcal{S}_2(x) + \frac{\pi}{4} \right], & \text{III: } x_0 + \delta_2 \le x \le x_0' - \delta_3, \\[12pt]
				\displaystyle
				(-1)^{p+1} \frac{x_0'^{1/3}}{(n n')^{1/12}} \, \frac{1}{x^{1/2}} \,
				\operatorname{Ai}\! \left( \frac{(n n')^{1/6}}{x_0^{2/3}} (x - x_0') \right), & \text{IV: } x_0' - \delta_3 < x < x_0' + \delta_4, \\[12pt]
				\displaystyle
				(-1)^{p+1} \frac{1}{\sqrt{2\pi}} \, \frac{1}{(x - x_0)^{1/4}(x - x_0')^{1/4}} \,
				\exp\!\left[ -\mathcal{S}_3(x) \right], & \text{V: } x \ge x_0' + \delta_4,
			\end{cases}
			\label{eq:I_WKB_full}
		\end{equation}
	\end{widetext}
	where $	\operatorname{Ai}(x)$ is an Airy function, $\mathcal{S}_1, \mathcal{S}_2, \mathcal{S}_3$ are the positive WKB action integrals in the classically allowed regions. The transitional widths $\delta_i$ scale as $\delta_1, \delta_2 \gtrsim x_0 / [(n n')^{1/6} (\sqrt{n}-\sqrt{n'})^{1/3}]$ and $\delta_3, \delta_4 \gtrsim x_0' / [(n n')^{1/6} (\sqrt{n}+\sqrt{n'})^{1/3}]$.
	
	Since $x \lesssim x_0$ in our problem, region II is the most relevant. In this region, the Airy function can be expressed in terms of modified Bessel functions, the approximation can be obtained
	\begin{align}
		I_{n,n'}(x) &\approx \frac{1}{\pi\sqrt{3}} \left(1 - \frac{x}{x_0}\right)^{1/2} K_{1/3}(z), \label{eq:I_approx_II} \\
		z &= \frac{2}{3} (x_0^2 n n')^{1/4} \left(1 - \frac{x}{x_0}\right)^{3/2}. \nonumber
	\end{align}
	
	The uniform WKB approximation for the radial functions \(I_{n,n'}(x)\) is constructed by matching asymptotic solutions across the turning points \(x_0\) and \(x_0'\). Here $p$ is an integer determined by the Bohr--Sommerfeld quantization condition:
	
	\begin{equation}
		\int_{x_0}^{x'_0} \sqrt{-f_{n,n'}(x')} \, dx' = \left( p - \frac{1}{2} \right) \pi .
		\label{eq:quantization}
	\end{equation}
	
	The integral in Eq.~\eqref{eq:quantization} can be evaluated exactly, but we only require the limit of large $n$ and $n'$, where
	\begin{equation}
		f_{n,n'}(x) \approx \frac{(x - x_0)(x - x'_0)}{4x^2}.
		\label{eq:approx}
	\end{equation}
	
	Performing the integration in Eq.~\eqref{eq:quantization} gives
	\begin{equation}
		\int_{x_0}^{x'_0} \sqrt{-f_{n,n'}(x')} \, dx' = \pi \left( \min\{n, n'\} + \frac{1}{2} \right).
		\label{eq:integral_result}
	\end{equation}
	
	Therefore,
	\begin{equation}
		p = \min\{n, n'\} + 1 .
		\label{eq:p_value}
	\end{equation}
	
	The relevant action integrals are
	\begin{align}
		\mathcal{S}_1(x) &= \int_x^{x_0} \sqrt{f(x')} \, dx', \quad x < x_0, \label{eq:S1_def} \\
		\mathcal{S}_2(x) &= \int_{x_0}^{x} \sqrt{-f(x')} \, dx', \quad x_0 < x < x_0', \label{eq:S2_def} \\
		\mathcal{S}_3(x) &= \int_{x_0'}^{x} \sqrt{f(x')} \, dx', \quad x > x_0'. \label{eq:S3_def}
	\end{align}
	Explicit expressions for these integrals in terms of elementary functions can be obtained but are omitted here for brevity. The related formulas yield the piecewise approximation Eq.~\eqref{eq:I_WKB_full} used in the main text.
	
	\section{Spin Transitions and Polarization Dynamics}
	\label{subsec:spin_transitons}
	\subsection{Spin transitions}
	\label{subsec:spin_polarization}
	
	The WKB approximation can be applied here to derive analytic expressions for the spin-flip transition rates. For large $n,  n'$, the combinations of $I$-functions appearing in the matrix elements Eqs.~\eqref{eq:alpha_12_elements} and \eqref{eq:alpha_3_element} are given in Eq.~\eqref{eq:I_K_relation}, which can be expressed linearly in terms of modified Bessel functions $K_{1/3}$ and $K_{2/3}$~\cite{SokolovTernov1986}. Combining all the relations, one can arrive at the compact result for the spin-dependent transition rate:
	\begin{multline}
		w(\zeta, \zeta') = \frac{5\sqrt{3}}{6} \frac{e_0^2}{\hbar c} \frac{c}{R} \frac{E}{m_0 c^2}
		\Biggl[ \frac{1+\zeta\zeta'}{2} \Bigl( 1 - \frac{16\sqrt{3}}{45} \xi_0 + \frac{25}{18} \xi_0^2 \\
		- \frac{\zeta'}{5} \xi_0 \bigl( 1 - \frac{20\sqrt{3}}{9} \xi \bigr) \Bigr)
		+ \frac{1-\zeta\zeta'}{2} \frac{\xi_0^2}{6} \bigl( 1 - \zeta' \frac{8\sqrt{3}}{15} \bigr) \Biggr].
		\label{eq:w_spin_final}
	\end{multline}
	The rate for a spin-flip transition ($\zeta' = -\zeta$) follows as
	\begin{equation}
		w_{\text{flip}}(\zeta) = \frac{1}{2\tau} \left( 1 + \frac{8\sqrt{3}}{15} \zeta \right), \frac{w_{\text{flip}}(-1\to 1)}{w_{\text{flip}}(1\to-1)}\approx0.0396,
		\label{eq:w_spin_flip}
	\end{equation}
	where the characteristic relaxation time $\tau$ is
	\begin{equation}
		\tau = \frac{8\hbar^2}{5\sqrt{3} m_0 c e_0^2}
		\left( \frac{m_0 c^2}{E} \right)^2
		\left( \frac{H_0}{H} \right)^3.
		\label{eq:tau_spin}
	\end{equation}
	Equation~\eqref{eq:w_spin_flip} reveals the asymmetry between spin-flip rates for $\zeta = +1$ and $\zeta = -1$, which is the origin of the Sokolov-Ternov effect leading to spontaneous spin polarization of the electron beam.

	\subsection{Spin polarization dynamics}
	\label{subsec:spin_polarization_dynamics-1}
	
	The time evolution of the spin populations is governed by the coupled rate equations
	\begin{align}
		\frac{dn_\downarrow}{dt} &= n_\uparrow w_{\text{flip},\zeta=1} - n_\downarrow w_{\text{flip},\zeta=-1}, \label{eq:spin_rate_eq-1-1} \\
		\frac{dn_\uparrow}{dt} &= n_\downarrow w_{\text{flip},\zeta=-1} - n_\uparrow w_{\text{flip},\zeta=1},
		\label{eq:spin_rate_eq-2-1}
	\end{align}
	where \(n_\downarrow\) and \(n_\uparrow\) denote the number of electrons with spin anti-parallel (\(\zeta=-1\)) and parallel (\(\zeta=+1\)) to the magnetic field, respectively. The total electron number is conserved:
	\begin{equation}
		n_\downarrow + n_\uparrow = n = \text{constant}.
		\label{eq:spin_conservation-1}
	\end{equation}
	For an initially unpolarized beam at \(t=0\), it gets
	\begin{equation}
		n_\downarrow(0) = n_\uparrow(0) = n/2.
		\label{eq:spin_initial-1}
	\end{equation}
	
	Using the spin-flip transition rates from Eq.~\eqref{eq:w_spin_flip}, the solution of Eqs.~\eqref{eq:spin_rate_eq-1-1}--\eqref{eq:spin_initial-1} is
	\begin{equation}
		\begin{pmatrix} n_\downarrow(t) \\ n_\uparrow(t) \end{pmatrix}
		= \frac{n}{30} \begin{pmatrix}
			15 + 8\sqrt{3}\left[1 - e^{-t/\tau}\right] \\
			15 - 8\sqrt{3}\left[1 - e^{-t/\tau}\right]
		\end{pmatrix},
		\label{eq:spin_solution-1}
	\end{equation}
	where the relaxation time \(\tau\) is given by Eq.~\eqref{eq:tau_spin}. The asymptotic polarization degree for \(t \gg \tau\) is defined as
	\begin{equation}
		\mathcal{P}_{\text{spin}} = \frac{n_\downarrow(\infty) - n_\uparrow(\infty)}
		{n_\downarrow(\infty) + n_\uparrow(\infty)}
		= \frac{\langle \mathfrak{s} \rangle}{\mathfrak{s}_{min}} \approx 0.9238,
		\label{eq:spin_polarization_degree-1}
	\end{equation}
    Here \(\langle \mathfrak{s} \rangle\) is the average spin of the electrons, and \(\mathfrak{s}_{\text{min}} = -1/2\) is its minimum value. The result corresponds to approximately 92.38\% of the net electrons having their spin antiparallel to the magnetic field. This stationary state polarization is illustrated in Fig.~\ref{fig:spin_dist}.
	
	Equivalently, the evolution of the average spin projection \(\langle \zeta \rangle\) can be derived from the transition rates. Using Eq.~\eqref{eq:w_spin_flip}, it can be obtained
	\begin{equation}
		\frac{d\langle \zeta \rangle}{dt} = \sum_{\zeta' = -\zeta} (\zeta' - \zeta) w(\zeta, \zeta')
		= -\frac{1}{\tau} \left( \langle \zeta \rangle + \frac{8\sqrt{3}}{15} \right),
		\label{eq:spin_avg_evolution-1}
	\end{equation}
	which integrates to
	\begin{equation}
		\langle \zeta(t) \rangle = -\frac{8\sqrt{3}}{15} \left[ 1 - e^{-t/\tau} \right].
		\label{eq:spin_avg_solution-1}
	\end{equation}
	Equation~\eqref{eq:spin_avg_solution-1} reduces to Eq.~\eqref{eq:spin_polarization_degree-1} in the equilibrium- state limit.
	
%%%%%%%%%%%%%%%%%%%%%%%%

	\end{document}